\documentclass{article}

\usepackage{graphicx}  
\usepackage{amsmath}   
\usepackage{amssymb}   
\usepackage{bm}         

\usepackage[compress]{cite}
\usepackage[T1]{fontenc}                
\usepackage{lipsum}
\usepackage{hyperref}
\usepackage{mathrsfs}

\usepackage{lscape}
\usepackage{longtable, supertabular}
\usepackage{times}

\usepackage{framed}

\def\eq#1{{Eq.~(\ref{#1})}}

\def\fig#1{{Fig.~\ref{#1}}}

\def\HI{{{\textrm{H}}I}}

\def\om{{\Omega_m}}

\def\oh{{H_0}}

\def\orr{{\Omega_R}}

  \title{\bf A Multi-messenger view of Cosmic Dawn:\\
\textit{Conquering the Final Frontier}}
\smallskip
\author{Hamsa Padmanabhan\footnote{\textit{Email:} hamsa.padmanabhan@unige.ch}\\
\\
Universit\'e de Gen\`eve, D\'epartement de Physique Th\'eorique,\\
24 quai Ernest-Ansermet, CH-1211 Gen\`eve 4, Switzerland\\
}

\date{}

\begin{document}
 
  \maketitle
  
  \hrule
 
  \begin{abstract}
The epoch of Cosmic Dawn, when the first stars and galaxies were born, is widely considered the final frontier of observational cosmology today. Mapping the period between Cosmic Dawn and the present-day provides access to more than 90\% of the baryonic (normal) matter in the Universe, and unlocks several thousand times more Fourier modes of information than available in today's cosmological surveys. We review the progress in modelling baryonic gas observations as tracers of the cosmological large-scale structure from Cosmic Dawn to the present day. We illustrate how the description of dark matter haloes can be extended to describe baryonic gas abundances and clustering. This innovative approach allows us to fully utilize our current knowledge of astrophysics to constrain cosmological parameters from future observations. Combined with the information content of multi-messenger probes, this will also elucidate the properties of the first supermassive black holes at Cosmic Dawn. We present a host of fascinating implications for constraining physics beyond the $\Lambda$CDM model, including tests of the theories of inflation and the cosmological principle, the effects of non-standard dark matter, and possible deviations from Einstein's general relativity on the largest scales.
 \end{abstract}
 \bigskip


\bigskip

\hrule

\newpage
\tableofcontents
\newpage

\section{Introduction}\label{s:intro}

The last few decades have witnessed several rapid advances in the field of observational cosmology, with  probes of the universe over a wide range of wavelengths from the radio to the X-ray bands, as well as — recently —in the gravitational wave regime. The wealth of observational data has led to the development of precision cosmology, whereby the cosmological parameters are measured to unprecedented accuracy with the available data. The fact that all the independent observational constraints are consistent with the `standard model' of theoretical cosmology, known as $\Lambda$CDM, is one of the most impressive successes of the theory. 

However, the standard model of cosmology also leaves us with several challenging problems. About 70\% of the present-day energy density of the universe is in the form of ‘dark energy’, which behaves like a smooth fluid having negative pressure, and leads to the accelerated expansion of the universe. This component is consistent with a cosmological constant term introduced in Einstein’s equations. Another 25\% is in the form of ‘dark matter’ — which does not interact with radiation, but participates in gravitational clustering. We do not have a physical understanding of (or laboratory evidence for) either of these two components, which together comprise about 95\% of the present-day universe. Only the remaining approximately 5\% of the energy density is in the form of ordinary matter (with a tiny amount in the form of radiation), whose physical properties are familiar to us. 

At the earliest epochs, radiation decoupled from the neutral baryonic matter in the first major phase transition of the observable universe known as the epoch of recombination, which occurred about 300,000 years after the Big Bang, and is observable today as the cosmic microwave background (CMB). Most of the baryonic material in the universe was (and is) in the diffuse plasma and gas between galaxies, known as the intergalactic medium (IGM) -- with a very small amount ($\sim 10\%$) in stars, as has been roughly the case throughout the timescale of the observable Universe \cite{shull2012, famaey2012}.\footnote{If it is sobering to note that we are not made up of the most abundant component of the Universe, namely dark energy, it is perhaps even more humbling to find that stars and stellar systems -- i.e., the baryonic material we are most familiar with -- did not even represent the majority of the cosmological \textit{baryons} over all epochs!}  The baryonic matter in the universe was almost fully hydrogen [with small ($\sim$ 10\%) amounts of neutral helium], in its electrically neutral, gaseous form (known as neutral hydrogen, hereafter referred to as HI, as commonly done in the literature) at the end of the epoch of recombination, and remained so (during a period known as the dark ages of the universe) until the first stars and galaxies formed about a few hundred million years later. These luminous sources contributed ionizing photons to complete the second major phase transition in the observable universe known as cosmic reionization. In this process, the radiation from starlight was primarily responsible for ionizing the hydrogen in the universe – and this period, which lasted for about a few hundred million years, that immediately followed the dark ages is referred to as the Cosmic Dawn. Mapping the period between Cosmic Dawn and the present-day provides access to more than 90\% of the Universe's baryonic (normal) matter, and unlocks almost all the information in cosmological baryons.

It is therefore obvious that our most accurate understanding of cosmology and fundamental physics in the future will come only through the study of the baryonic gas tracers of the Universe, the majority of which lie between the local universe probed by galaxy surveys, and the CMB surface of last scattering. This allows access to almost three orders of magnitude more independent Fourier modes of information than currently available from the combination of galaxy surveys (which extend only to the edge of the local universe), and the CMB.  In the last several years, a tremendous effort to revolutionise our understanding of cosmology from baryonic gas is bearing fruit. Studies of neutral hydrogen in the intergalactic medium at the epoch of Reionization and Cosmic Dawn are chiefly probed by the 21 cm spin-flip transition in the radio band. After reionization, neutral hydrogen exists in the form of dense clumps in galaxies, and extremely dense systems known as Lyman-limit systems and Damped Lyman-Alpha systems (DLAs). The carbon monoxide (CO) molecular abundance also offers exciting prospects for placing constraints on the global star-formation rate, which is observed to peak around 2 billion years after the Big Bang \cite{lilly1996, madau1998}. The technique of intensity mapping (IM) has emerged as the powerful tool to explore this phase of the Universe by measuring the integrated emission from sources over a broad range of frequencies, providing a tomographic, or three-dimensional picture of the Universe. Notably, this is being complemented by large scale efforts to probe the first galaxies and black holes by using the combined power of the electromagnetic and gravitational wave bands. We are thus on the threshold of the richest available cosmological dataset in the coming years, facilitating the most precise constraints on theories of Fundamental Physics. 

This review will address the major milestones in the field of cosmology and astrophysics with baryonic tracers. We begin (in Sec. \ref{sec:astrocosmo}) with an overview of the astrophysical and cosmological background needed for the results (readers familiar with cosmology and structure formation could skip Sec. \ref{sec:introcosmo}). In the next section, we provide a detailed overview of state-of-the-art techniques for modelling neutral gas occupation in dark matter haloes, with emphasis on hydrogen [both in galaxies, Sec. \ref{sec:21cmgal}, and in Damped Lyman Alpha systems (DLAs), Sec. \ref{sec:dlahimodels}]. We then (Sec. \ref{s:submmhalomodel}) illustrate how this framework can be carried over to the sub-millimetre regime, exploring the carbon monoxide (CO) and singly ionized carbon [CII] line transitions. We then describe how these models can be combined with the latest available data to constrain the best-fitting values of the parameters (Sec. \ref{s:constraints}) and subsequently can be used to (Sec. \ref{s:forecasts})  forecast constraints on cosmology and astrophysics, particularly on physics beyond the standard $\Lambda$CDM cosmological model. Finally, we describe (Sec. \ref{sec:crosscorrelations}) how the different probes of baryonic tracers can be effectively combined to produce cross-correlation forecasts, significantly increasing the sensitivity of the detections and offering a holistic view into the epoch of Cosmic Dawn.  We summarize the theoretical and observational outlook, and discuss open problems and future challenges in a concluding section (Sec. \ref{sec:outlook}). 

\section{Overview of cosmology and structure formation}
\label{sec:astrocosmo}
In this section, we provide a review of cosmology, structure formation and the key milestones in the evolution of baryonic material. We also discuss the major observational probes of gas within and around galaxies.
\subsection{The cosmological background}
\label{sec:introcosmo}

Current observations indicate that our universe is homogeneous and isotropic on the largest scales, to the level of better than 1 part in $10^5$. 
According to the standard model of cosmology (the hot Big Bang model), the homogeneous and isotropic universe is described by the Friedmann-Robertson-Walker (FRW) metric of the form:
\begin{equation}
ds^2 = -c^2 dt^2 + a(t)^2 \left[\frac{dr^2}{1 - Kr^2} + r^2(d \theta^2 + \sin^2 \theta d \phi^2)\right]                                                                                  \end{equation} 
where $K$ is a constant which determines the curvature and can take the values $0, 1, -1$, $a(t)$ is called the cosmic scale factor, and $r, \theta, \phi$ are comoving spherical coordinates centered on the observer. The Einstein field equation relates the geometry of this spacetime to the underlying matter-energy content (including the vacuum energy, taken to be a cosmological constant term $\Lambda$ throughout this review):
\begin{equation}
 R_{ij} - \frac{1}{2} g_{ij} R + g_{ij} \Lambda = \frac{8 \pi G}{c^4} T_{ij}
\end{equation} 
In the above equation, $R_{ij}$ is the Ricci curvature tensor of the spacetime, $g_{ij}$ is the metric, $R$ is the Ricci scalar,  $T_{ij}$ is the energy-momentum tensor of the matter content of the universe, which has the form $T^i_j = {\rm diag}(-\rho c^2, p, p, p)$ with $\rho c^2$ being the energy density and $p$ being the pressure.  With this, 
 the Einstein equation for the FRW metric can be shown to have the form:
 \begin{equation}
 \left( \frac{\dot{a}}{a} \right)^2 = \frac{8 \pi G}{3} \rho - \frac{K c^2}{a^2} + \frac{\Lambda c^2}{3}
 \end{equation} 
 The above equation is known as the Friedmann equation. 
 Throughout this review,  we will work with the flat universe in which the curvature component $K = 0$.
 A source located at the physical distance $l = a(t) r$ from an observer moves at a velocity  $v = dl/dt = \dot{a} l = \dot{a} r/a(t)$ (due to the expansion of the universe, assuming that the source does not have a peculiar velocity of its own). If we define 
 the Hubble parameter $H(t) = \dot{a} (t)/ a(t)$ at the cosmic time $t$, we  obtain the relation $v = H(t) r$, which is known as Hubble's law. Radiation emitted by a source at time $t$ is observed today (at time $t_0$) with a redshift $z = a(t_0)/a(t) - 1$.
 The radiation and matter content of the universe are usually parametrized in terms of the dimensionless numbers, $\orr$ and $\om$, which are defined as
\begin{equation}
 \orr \equiv \frac{8\pi G \rho_R(t_0)}{3 \oh^2}; \quad \om \equiv \frac{8\pi G \rho_m(t_0)}{3 \oh^2}
\end{equation} 
Here $t_0$ is the current age of the universe, $H_0\equiv(\dot a/a)_{t_0}$ is the present value of the Hubble parameter, and $\rho_R(t_0), \rho_m(t_0)$ are the energy densities of radiation and matter in the universe at $t= t_0$. 
In terms of these components of the universe, the Friedmann equation can be expressed as:
\begin{equation}
 \frac{\dot a^2}{a^2} = H_0^2 \left[ ( 1 - \Omega_R - \Omega_m)+ \frac{\Omega_R a_0^4}{a^4} + \frac{\Omega_m a_0^3}{a^3}\right]
\label{a1}
\end{equation} 
where the symbols $\orr$ and $\om$ refer to density parameters at the present epoch. The present value of the Hubble parameter, $H_0$, is usually specified in terms of the dimensionless number $h$, as $H_0 = 100 h$ km s$^{-1} {\rm Mpc}^{-1}$. 

The earliest epoch of the universe consisted of an inflationary phase characterized by rapid exponential expansion; subsequently, the universe entered the radiation dominated phase. The quantum-mechanical vacuum fluctuations during inflation generated small inhomogeneities that manifested as perturbations in the matter-radiation components of the universe.  The formation of light nuclei occurred when the universe had cooled to a temperature $T_{\rm nuc} \sim 0.1$ MeV. The contributions to the baryonic mass were roughly $\sim 75\%$ hydrogen and $25\%$ helium nuclei, with very small amounts of deuterium, lithium and other light elements.  The transition from the radiation to matter (consisting mostly of dark matter) dominated epoch occurred around redshift $z \sim 3300$. Around $z \sim 1100$, the electrons and protons recombined to form neutral atoms, when the temperature was around $T_{\rm rec} \sim 0.3$ eV. This was also the epoch at which the ambient radiation decoupled from the baryonic matter.
The primordial density perturbations, which had the fractional amplitude of $\sim 10^{-5}$ grew over time in the matter dominated epoch of the universe. The perturbations collapsed to form gravitational bound objects known as haloes, which were initially formed on small scales. 
In the simplest model of spherical collapse, an overdense region expands more slowly as compared to the background, reaches a maximum radius, turns around and contracts and virializes to form the bound halo.
In a flat matter-dominated universe,
this model leads to the result that the density of the collapsed halo is about a factor $18 \pi^2 \approx 178$ times the background density at the time of collapse. 
 In a universe with $\om + \Omega_{\Lambda} = 1$, the expression for the overdensity of the virialized structure is well-described by the fitting formula:
 \begin{equation}
  \Delta_v \approx 18 \pi^2 + 82 d - 39 d^2
 \end{equation} 
 where $d = \Omega_m(z) - 1$, with \cite{bryan1998}:
 \begin{equation}
  \Omega_m(z) - 1 = \Omega_m (1+z)^3/(\Omega_m (1+z)^3 + \Omega_{\Lambda}) - 1
 \end{equation} 
 By solving the equation of motion of collapse and using the virial theorem,  we can obtain the characteristic radius and virial velocity of the collapsed structures, and relate them to the mass and the redshift of collapse.
 Halos of mass $M$ collapsing at redshift $z (\gg 1)$ are found to have the virial radius:
   \begin{equation}
 R_v (M) = 46.1 \ {\rm{kpc}} \  \left(\frac{\Delta_v \Omega_m h^2}{24.4} \right)^{-1/3} \left(\frac{1+z}{3.3} \right)^{-1} \left(\frac{M}{10^{11} M_{\odot}} \right)^{1/3}
 \label{virialradius}
 \end{equation} 
 and the corresponding virial (or circular) velocity of the halo:
 \begin{equation}
  v_c = \left(\frac{G M}{R_v(M)}\right)^{1/2}
  \label{virvel}
 \end{equation} 
 Even though the collapse of an overdense region is characterized by nonlinear gravitational growth, it is also useful in several contexts to define the quantity \textit{linearly extrapolated overdensity} $\delta_c(z)$ at every redshift. This corresponds to the overdensity predicted by linear theory and extrapolated into the nonlinear epoch. This linearized overdensity, extrapolated to the present, of an collapsed object at redshift $z$ (in the matter-dominated era) can be shown to be given by $\delta_c(z) \approx 1.686(1 + z)$.
 The theories of structure formation also predict the abundance of dark matter haloes i.e. the number density of haloes as a function of mass, at any redshift. The Press-Schechter model, based on the idea of spherical collapse, predicts the comoving number density of dark matter haloes to be given by:
 \begin{equation}
  n(M) = \frac{\rho_m}{M} \frac{d \ln \sigma^{-1}}{d M} f(\sigma)
 \end{equation} 
 where $\sigma(M)$ is the variance of the initial density field (assumed to be a Gaussian random field) on mass scale $M$, and 
 \begin{equation}
f(\sigma) = \sqrt{\frac{2}{\pi}} \frac{\delta_c(z)}{\sigma} \exp\left(-\frac{\delta_c^2}{2 \sigma^2} \right)
\end{equation}  
 where $\delta_c(z)$ is the linearly extrapolated overdensity at the collapse of the halo. 
 Detailed numerical calculations indicate that ellipsoidal, instead of spherical collapse, may be a better description to the abundance of haloes and has led to the Sheth-Tormen (ST) mass function \cite{sheth2002}:
 \begin{equation}
 f(\sigma) = A \sqrt{\frac{2 a}{\pi}} \frac{\delta_c(z)}{\sigma} \exp\left(-\frac{a \delta_c^2}{2 \sigma^2} \right) \left[1 + \left(\frac{\sigma^2}{a \delta_c^2}\right)^p \right]
 \end{equation} 
 where $a = 0.707, p = 0.3$, and $A = 0.322$. 
 
 The clustering of the dark matter haloes 
 reflects the effects of gravitational growth and
 describes the deviation from the smooth universe. This clustering is well described by the bias function $b(M,z)$ which specifies the clustering of the dark matter haloes as a function of halo mass.
 In the Press-Schechter model,  the bias function takes the form:
 \begin{equation}
b_{\rm PS}(M)  = 1 + \frac{\nu_c^2 - 1}{\delta_{c} (z = 0)}   
\end{equation} 
where $\nu_c = \delta_c(z)/\sigma(M)$. In the Sheth-Tormen formalism, the corresponding expression is given by \cite{scoccimarro2001}:
\begin{equation}
 b_{\rm ST}(M)  = 1 + \frac{a \nu_c^2 - 1}{\delta_{c} (z = 0)} + \frac{2 q/\delta_{c} (z = 0)}{1 + (a \nu_c^2)^q}
\end{equation} 
It is to be noted that the very small haloes can become ``antibiased'' $(b_{\rm ST} < 1)$ at late times, because these tend to form in underdense regions (the overdense regions being already taken up by more massive haloes).

To a good approximation, the dark matter haloes may be modelled as spherical objects. Numerical simulations indicate a near-uniform radial density profile $\rho(r)$ for the radial structure of the dark matter haloes. One of the widely used forms is known as the Navarro-Frenk-White (NFW) profile. In this model, the dark matter haloes are described as spheres with the density distribution:
\begin{equation}
 \rho(r) = \frac{\rho_0}{(r/r_s)(1 + (r/r_s))^2} 
\end{equation} 
where $r_s$ is known as the scale radius, and $\rho_0$ is a characteristic density. This profile is found to be a good model of dark matter haloes over a range of masses in cold dark matter cosmologies. The scale radius is related to the virial radius defined in \eq{virialradius}, by $r_s = R_v(M)/c(M,z)$ where the concentration parameter, $c(M,z)$ is given by:
\begin{equation}
    c(M,z)  = \frac{9}{1+z} \left(\frac{M}{M_*(z)}\right)^{-0.13}
    \label{concparamdm}
\end{equation}
and $M_*$ is defined as the halo mass at which $\nu_c \equiv \delta_c(z)/\sigma(M) = 1$.

The \textit{halo model} of dark matter clustering unifies the ingredients of the halo distribution above (the mass function, clustering, and density profiles) into a framework for computing the statistical properties of the collapsed objects even into the nonlinear regime. The model is found to work extremely well when compared to the results of numerical simulations. In the halo model, the parameter $\rho_0$ in the profile is fixed by normalizing the total mass of dark matter in the halo:
\begin{equation}
    M = 4 \pi \int_0^{R_{v}} \rho(r) dr 
\end{equation}
where it is assumed that the halo is truncated at its virial radius. Given this parametrization, the clustering of the haloes is computed by defining the Fourier transform of the profile (normalized to unity on large scales):
\begin{equation}
    u(k|M) = \frac{4 \pi}{M} \int_0^{R_v} dr \ r^2 \frac{\sin kr}{kr}{\rho(r)}
    \label{udm}
\end{equation}
in terms of which the \textit{power spectrum} of density fluctuations can be defined as:
\begin{equation}
    P(k,z) = P_{1h} (k,z) + P_{2h} (k,z)
\end{equation}
having two terms, the \textit{two-halo} term, $P_{2h} (k,z)$ describing correlations between particles in different haloes, and the \textit{one-halo} term, $P_{\rm 1h}(k,z)$ describing the structures within each halo:
\begin{eqnarray}
    P_{1h}(k, z) &=& \int dM \  n(M, z) \left(\frac{M}{\bar{\rho}}\right)^2 |u(k|M)|^2 \nonumber \\
     P_{2h}(k, z) &=& P_{\rm lin} (k) \int dM \ \left[ \left(\frac{M}{\bar{\rho}}\right) b(M,z)  n(M, z)  |u(k|M)|\right]^2
     \label{powerspecdm}
\end{eqnarray}
where $P_{\rm lin}$ is the power spectrum computed  in the linear theory, and $\bar{\rho}$ is the mean dark matter density, defined as:
\begin{equation}
    \bar{\rho} = \int dM n(M,z) M 
\end{equation}

\subsection{Evolution of baryons}
The baryons in the very early universe were in the form of a hot, ionized plasma which was tightly coupled to the ambient radiation. At a redshift of $z_{\rm rec} \sim 1100$,  when the universe had expanded and cooled enough, the protons and electrons in the plasma "re"-combined to form neutral hydrogen atoms. At this stage, the ambient radiation decoupled from the newly formed neutral atoms.
After the recombination epoch, the gas in the universe was predominantly neutral (to better than 1 part in $10^4$) and the Universe entered the ``dark ages'' which were characterized by the absence of radiation (other than the relic CMB radiation). For the first hundreds of millions of years after the recombination, the baryons in the Universe were mostly in the form of neutral hydrogen ($\sim 93\%$ by number) and small amounts of neutral helium ($\sim 7\%$ by number), with negligible amounts of heavier elements.

As the dark matter haloes formed by gravitational clustering, the baryonic material (hydrogen, with small amounts of helium) was pulled into the potential wells created by the dark matter haloes. The speed of infalling gas is of the order of the virial velocity $v_c$ of the halo, \eq{virvel}. As the gas cools and becomes Jeans unstable, i.e. the total gas mass exceeds the Jeans mass, it is able to form stars.  The first stars thus formed in the universe were massive, luminous and metal-free,  and termed as Population III stars. In the $\Lambda$ CDM models, these stars had masses $\sim 10^5 M_{\odot}$ and formed at $z \gtrsim 30$. However, the majority of the matter today -- and at the early epochs of first stars -- was outside these structures, and resided in the intergalactic medium (IGM) -- which is the term used to describe the baryonic material lying between galaxies and not directly associated with dark matter haloes.  At sufficiently early epochs, of course, all the baryons in the universe resided in the IGM. Today, only about 10\% of the baryons are locked up in luminous structures and the remaining are believed to reside in the IGM (with a small fraction in the intra-cluster medium of galaxies and the circumgalactic medium surrounding galaxies). The IGM provides the fuel for star and galaxy formation, interacts with the radiation emitted by the galaxies, and is enriched by the gas and metal outflows from galaxies. Being fairly less affected by the nonlinear astrophysical processes associated with galaxy formation, it offers a much clearer view of the underlying cosmology, the cosmological changes since recombination, and the physics of structure formation and clustering. 

As soon as the first luminous sources formed, they started to emit UV radiation and ionize the hydrogen gas in the IGM back into free electrons and protons -- this process is referred to as cosmological reionization. Both the temperature and the ionization state of the gas were affected by the reionization process. The epoch of reionization, which was the epoch at which the majority of hydrogen got ionized, thus represents the second major phase transition (the first being recombination) of the baryonic material in the universe.

\subsection{Cosmic reionization}
\label{sec:reion}

The epoch of reionization is associated with the end of the ``dark ages'' of the universe. The majority of the baryonic material in the universe is hydrogen, and hence ``reionization'' usually refers to the process of hydrogen reionization.  Various probes are used to determine the epoch of hydrogen reionization in the universe (for reviews on reionization, we refer to Refs. \cite{loeb2001, barkana, lidz2016rev, fan2006, choudhury2006, choudhury2009, zaroubi2013, natarajan2014, ferrara2014}):

(a) The Thomson scattering optical depth of the cosmic microwave background offers an integrated probe of the free electron density between the present epoch and the epoch of reionization. If reionization is assumed to be a sharp transition from fully neutral to fully ionized IGM, occurring at redshift $z_{\rm ri}$, the expression for the Thomson optical depth along a line-of-sight is:
\begin{equation}
 \tau_{\rm Th} = \int_0^{z_{\rm ri}} n_e(z) \ \sigma_e \  dz \left|\frac{dl(z)}{dz}\right|
\end{equation} 
For the standard $\Lambda$CDM cosmology, the above integral can be done analytically and leads to an expression in terms of the standard cosmological parameters and the primordial helium abundance \cite{venkatesan2008}. In the high-redshift limit of $\Omega_m (1+z)^3 \gg \Omega_{\Lambda}$, the expression simplifies to \cite{moandwhite}:
\begin{equation}
 \tau_{\rm Th} = 0.07 \left(\frac{h}{0.7}\right) \left(\frac{\Omega_{b}}{0.04}\right) \left(\frac{\Omega_{m}}{0.3}\right)^{-1/2} \left(\frac{1 + z_{\rm ri}}{10}\right)^{3/2}
\end{equation} 
where $\sigma_e$ is the Thomson scattering cross-section and $n_e(z)$ is the comoving number density of electrons at redshift $z$ along the sightline. 
 The WMAP temperature and polarization measurements of optical depth ($\tau_e = 0.087 \pm 0.017$) pointed to the (instantaneous) redshift of reionization being about $z_{\rm ri} \sim 11.0 \pm 1.4$ \cite{dunkley2009}.  The recent Planck measurements \cite{planck2015}, on the other hand, favour a later epoch of completion of reionization ($z \sim 6-10$, Ref. \cite{mitra2019, robertson2015}) due to the smaller measured value of the Thomson optical depth ($\tau_e = 0.066 \pm  0.016$). In reality, reionization is not a sharp transition and hence the optical depth measurement only provides an estimate of the average reionization epoch.
 
(b) A second probe of the epoch of reionization is the Gunn-Peterson optical depth of redshifted Lyman$-\alpha$ photons in the spectra of high-redshift quasars.\footnote{ {Strictly speaking, the term `quasar' is commonly used  for describing radio-loud quasi-stellar objects. However, as frequently done in the literature,  we will use the terms `quasars' or QSOs interchangeably to indicate quasi-stellar objects throughout this review, irrespective of their radio properties.}}
 
 The hydrogen Lyman-$\alpha$ line, which denotes the transition from the $n = 1$ to $n = 2$ electronic states, has been a powerful probe of hydrogen in cosmology, in the spectra of astrophysical objects such as quasars, galaxies and gamma-ray bursts (GRBs). The rest wavelength of the line is $\lambda_{\alpha} = 1216$ \AA. Photons from a cosmological source, whose wavelength is less than 1216 \AA\ are redshifted into the Lyman-$\alpha$ wavelength as they propagate through the IGM. These can be absorbed by a hydrogen atom and re-emitted in a different direction. The optical depth of the IGM can thus be probed by the flux decrement in the spectra of high-redshift luminous sources. 

The optical depth from absorption of redshifted Lyman-$\alpha$ photons (with $\Lambda_{\alpha}$ being the decay rate) from a uniform neutral hydrogen distribution is given by \cite{gunnpeterson}: 
\begin{eqnarray}
 \tau(z) &=& \frac{3 \Lambda_{\alpha} \lambda_{\alpha}^3 x_{\rm HI} n_{\rm H} (z)}{8 \pi H(z)} \nonumber \\
 &\sim & 1.6 \times 10^5 x_{\rm HI} (1 + \delta) \left(\frac{1+z}{4}\right)^{3/2}
\end{eqnarray} 
assuming standard values for the cosmological parameters, the Hubble parameter $H(z)$ evaluated during the matter-dominated era, and $n_{\rm H}(z) = \bar{n}_{\rm H}(z) (1 + \delta)$ where $\bar{n}_{\rm H}$ is the mean cosmic hydrogen density. 
Here, $x_{\rm HI}$ is the neutral fraction of hydrogen. We thus see that a neutral fraction of even about 1 part in $10^4$ (i.e. $x_{\rm HI} \sim 10^{-4}$) would give rise to complete Lyman-$\alpha$ absorption. Any transmission, therefore, implies a highly ionized (better than 1 part in $10^4$) IGM. At moderate redshifts ($z < 5.5$), \textit{no  complete absorption signatures} (known as the Gunn-Peterson absorption troughs) are seen in the spectra of quasars. Hence, it is concluded that the IGM must be very highly (re)ionized upto these redshifts. 

Recent results based on the spectra of quasars above $z \sim 6$ show the first hints of Gunn-Peterson absorption troughs indicating the possible transition to the neutral IGM. However, because of the small neutral fraction required to saturate the Gunn-Peterson trough, it is difficult to place constraints on the level of ionization of the IGM at these redshifts. Hence, this data alone does not constrain the history of reionization, i.e. whether reionization was a sharp transition occurring around redshift 6 or a fairly gradual transition beginning from a much higher redshift.  The latest studies combining the CMB measurement with the spatial fluctuations in the IGM optical depth \cite{kulkarni2019, nasir2020, raste2021, davies2016} may indicate a late, rapid end to reionization occurring around $z \sim 5.3$, however, these fluctuations could also arise due to variations in mean free paths of hydrogen atoms in an  ultraviolet background dominated by galaxies \cite{davies2016}, or due temperature fluctuations in the IGM \cite{dalonsio2015} in more extended reionization scenarios ending at $z \sim 6$.

(c) The hydrogen 21-cm spin-flip line is a third important probe of the cosmic reionization. This line is inherently weak, and hence does not get completely saturated even over the early-to-mid stages of reionization. Being a line transition, it also provides a three-dimensional probe of the universe (two dimensional surface information at every frequency). Since the sensitivity extends to the Jeans' scale of the baryonic material, it is unaffected by Silk damping and hence gives access to many more modes than the CMB. The 21-cm transition is described in further detail in Sec. \ref{sec:21cmintro}.

Apart from the probes described above, studies of Lyman-$\alpha$ emitters (LAEs) (for a review, see Ref. \cite{ouchi2019}) and $\gamma$-ray bursts (GRBs), e.g., \cite{totani2006, kistler2009, ishida2011,robertsonellis2012, lidz2021a} and Lyman-Break Galaxies (LBGs, Ref. \cite{kashino2020}) are also useful probes of reionization. Studies of the high-redshift LAE and LBG data may also favour a late reionization scenario, e.g., \cite{choudhury2015, mesinger2015, kashino2020}.

\subsection{The first black holes}
\label{sec:firstbh}

The period of Cosmic Dawn also overlapped with the appearance of the first black holes, whose formation mechanism is considered one of the most outstanding challenges in cosmology today. Hence, gaining insights into their properties by means of their signatures on the environment is a complementary tool to the study of the intergalactic medium. The black hole masses are related to host halo mass through the black hole mass - bulge mass relation; (for a review, see, e.g.,  \cite{kormendy2013}).
The circular velocity of the dark matter halo, $v_{c}$, is found to tightly correlate with the velocity dispersion of the galaxy, which can then be used to relate the mass of the black hole to that of the halo, through:
\begin{equation}
    M_{\rm BH} \propto v_c^{\gamma} 
\end{equation}
with $\gamma \sim 4-5$, e.g., \cite{wyithe2002}. For the host galaxies, data-driven formalisms, e.g., \cite{behroozi2010} find the stellar mass to evolve as a broken power law with respect to the host halo mass:  $M_* \propto M^{2.3}$
when $M \leq 10^{12} M_{\odot}$, and 
$M_* \propto M^{0.29}$
for $M
\geq 10^{12} M_{\odot}$. These treatments can thus be used to relate the black hole mass to the host halo mass by substituting for the circular velocity from \eq{virvel} and introducing an overall normalization factor, $\epsilon_0$ \cite{wyithe2002}:
\begin{equation}
    M_{\rm BH} (M) = M \epsilon_0 \displaystyle{\left(\frac{M}{10^{12} M_{\odot}}\right)^{\gamma/3 - 
1}} \left(\frac{\Delta_v \Omega_m h^2}{18 \pi^2}\right)^{\gamma/6} 
(1+z)^{\gamma/2} 
\end{equation}

Massive black holes in binary systems during the epoch of reionization may be detected via the gravitational waves that they emit in the milliHertz to nanoHertz regime, with upcoming facilities such as the Pulsar Timing Arrays (PTAs) and the Laser Interferometer Space Antenna (LISA).  This allows the inference of several properties of the black holes, such as the masses of the binary members, the rough sky location and the source luminosity distance (which can be converted into an equivalent redshift for an assumed cosmology). This information complements the effect of the black hole and its host galaxy (or quasar) on the intergalactic medium, such as the ionization of the gas, and hence can be used as a powerful complementary probe of the evolution of reionization.

\subsection{The intergalactic medium : quasar absorption lines}
\label{sec:igm}

As we have seen, the optical depth of the IGM to photons from quasars provide useful clues to the timing and average redshift of reionization. The absorption lines in the spectra of quasars are also effective probes of the physical state of the IGM. The evolution of the IGM over a large timescale can be studied with the quasar absorption lines, since the quasars are observed out to fairly high redshifts, $0 < z < 5$.
 Neutral hydrogen clouds, that occur at different positions along the line-of-sight from the observer to the distant quasar, absorb Lyman-$\alpha$ photons at frequencies corresponding to the redshifts of the individual clouds. 

The absorption line profiles are expressed using the Voigt function which accounts for both the Doppler (thermal, with some turbulent) broadening which is dominant in the central region of the line, and the natural broadening which is dominant in the wings (for a review, see \cite{djikstra2014}). The Voigt profile is a convolution of these two mechanisms:
\begin{equation}
 \phi(\nu) = \int_{-\infty}^{\infty} P(v) L[\nu(1 - v/c)] \ d v
\end{equation} 
where $P(v)$ is the Maxwellian distribution of the velocities of the absorbing atoms:
\begin{equation}
P(v) \ dv = \frac{1}{\sqrt{\pi b^2}} \exp\left(-\frac{v^2}{b^2}\right) \ dv
\end{equation} 
and $b$ is the Doppler parameter of the gas and includes contributions of both thermal and turbulent motions:
\begin{equation}
 b^2 = \frac{2 k_B T}{m} + b_{\rm turb}^2
\end{equation} 
The line profile, $L$, occurs due to the natural line broadening in accordance with the Heisenberg uncertainty principle and the finite lifetime of the excited state. It has the form of a Lorentzian centered around the frequency of the transition $\nu_{\alpha}$:
\begin{equation}
 L(\nu) = \frac{1}{\pi} \frac{\gamma}{(\nu - \nu_{\alpha})^2 + \gamma^2}
\end{equation} 
where $\gamma = A_{21}/4\pi$, $A_{21}$ being the spontaneous transition coefficient. 

The Voigt profile can thus be expressed as:
\begin{equation}
 \phi(\nu) = \frac{1}{\sqrt{\pi}} \frac{c}{b} \frac{V(\nu)}{\nu}
\end{equation}
where
\begin{equation}
V(\nu)  = \frac{A}{\pi}  \int_{-\infty}^{\infty} dy \ \frac{e^{-y^2}}{(B - y)^2 + A^2} 
\end{equation}
with $A = c\gamma/b \nu$; $B = c(\nu - \nu_{\alpha})/b \nu$.
The Voigt profile can be approximated to lowest order as:
\begin{equation}
 V(\nu) \approx \exp(-B^2) + \frac{1}{\sqrt{\pi}} \frac{A}{A^2 + B^2}
\end{equation} 
The above expression clearly separates the thermal broadening (near the center of the profile) and natural broadening (near the wings of the profile). 

The above discussion was for the case of a single absorption line profile. The numerous neutral hydrogen clouds along the line-of-sight to a luminous source (such as a quasar) typical produce their own line profiles.
The collective set of absorption lines produced by the neutral hydrogen clouds from the low-to moderate density IGM (with typical overdensities $\Delta \lesssim 10$) along the line-of-sight to a luminous source such as a quasar,  is known as the Lyman$-\alpha$ forest. 
The forest is believed to arise from the baryonic fluctuations during hierarchical clustering \cite{cen, zhang, miralda, hernquist}, and hence is a probe of the sheets, filaments and voids in the cosmic web. After reionization, an ionizing  background radiation is established, and the gas in the Lyman$-\alpha$ forest is in photoionization equilibrium with the background. 

The column densities (i.e., number densities measured along the column of the line-of-sight) of the Lyman-$\alpha$ forest lines are in the range $10^{12} - 10^{17}$ cm$^{-2}$. 
Clouds having column densities higher than this are known as Lyman-limit systems (LLSs), which are optically thick to the ionizing radiation.  If the column densities are above $10^{20.3}$ cm$^{-2}$, the systems have prominent damping wings from the natural line broadening, and are therefore classified as Damped Lyman Alpha systems (DLAs). Absorption systems having column densities between $10^{19}$ to $10^{20.3}$ cm$^{-2}$ are termed as sub-Damped Lyman Alpha systems (sub-DLAs).

The number density of IGM absorbers in a column density interval $(N_{\rm HI}, N_{\rm HI} + d N_{\rm HI})$ and in a redshift interval $(z, z+dz)$ is given by  ${d^2 \mathcal{N}/d N_{\rm HI} dz}$, where $\mathcal{N}$ is the observed number of absorbers. A quantity known as the column density distribution is frequently used in observational studies, denoted as:
\begin{equation}
 f_{\rm HI} (N_{\rm HI}, X) = \frac{d^2 \mathcal{N}}{d N_{\rm HI} dX} 
 \label{coldens}
\end{equation} 
measured with respect to the interval $X$, defined through $dX/dz = H_0 (1+z)^2/H(z)$.

Once the distribution function is known, the comoving density of neutral hydrogen may be computed as:
\begin{equation}
 \rho_{\rm HI} (z) = \frac{m_H H_0}{c} \int f_{\rm HI}(N_{\rm HI}, z) N_{\rm HI} d N_{\rm HI}
\end{equation} 
where $m_H$ is the hydrogen mass,
and this is typically expressed in terms of the critical density at $z = 0$:
\begin{equation}
 \Omega_{\rm HI} (z) = \frac{\rho_{\rm HI} (z)}{\rho_{c,0}}
\end{equation}

\subsection{The 21-cm transition}
\label{sec:21cmintro}

As we have seen above, absorption of the Lyman-$\alpha$ radiation against a luminous source (such as a bright quasar or a bright GRB afterglow at high redshifts) allows us to probe the number density of neutral hydrogen atoms. The Lyman$-\alpha$ transition is therefore an excellent probe of the low column density, partially ionized gas in the post-reionization universe. However, as discussed in Sec. \ref{sec:reion}, due to the large optical depth of the Lyman-$\alpha$ absorption (the Einstein A-coefficient is very large, $A_{{\rm{Ly}} \alpha} \sim 10^{7}$ s$^{-1}$) the IGM is opaque to Lyman-$\alpha$ absorption even if its neutral fraction is as small as $f_{\rm HI} \sim 10^{-4}$, which happens above $z \sim 6$. 

Using the 21-cm transition of neutral hydrogen provides complementary information about the evolution of the baryonic material. This transition takes place between the two hyperfine states corresponding to parallel and antiparallel spins of the proton and electron in the hydrogen atom. 
In contrast to the Lyman-$\alpha$ transition, the 21-cm line is strongly forbidden (the Einstein A-coefficient is $A_{10} = 2.85 \times 10^{-15} s^{-1}$ which corresponds to a lifetime $\sim 10^8$ years). Hence, the line is inherently weak, and difficult to observe in the laboratory. However, it is used extensively in  astrophysical contexts to study the hydrogen gas in and around the Milky Way and other galaxies. The inherent weakness of the line transition also prevents the saturation of the line, thus enabling it to serve as a direct probe of the neutral gas content of the intergalactic medium during the dark 
ages and cosmic dawn prior to hydrogen reionization. At later epochs, over the last 12 billion years (redshifts 0 to 5), the 21-cm line emission of HI is expected to trace the underlying dark matter distribution, due to the absence of the complicated reionization astrophysics \cite{bharadwaj2001a, bharadwaj2001, bharadwaj2004, wyithe2008, wyithe2009, bharadwaj2009, wyithe2010} and since most of the neutral hydrogen at these redshifts ($z \sim 2-5$) resides in the high column-density systems which are preferentially probed by this line. 

 The 21-cm emission line also allows for the measurement of the intensity of fluctuations across frequency ranges or equivalently across cosmic time, thus making it a three-dimensional, or tomographic, probe of the universe. This is because at every frequency interval, one has access to a two-dimensional surface. This 2D surface, along with the frequency (or equivalently, redshift) being the third ``axis'', gives us the three-dimensional information. The combination of angular and frequency structure allows us to use the 21-cm line for  tomography, or mapping almost 90\% of the baryonic material from Cosmic Dawn to the present day (e.g.,\cite{loeb2013}). This is a much larger comoving volume than galaxy surveys in the visible band, and consequently promises a much higher precision in the measurement of the matter power spectrum and cosmological parameters. Since the power spectrum can be probed to the Jeans' length of the baryonic material, it allows a sensitivity to much smaller scales than those allowed by the CMB.

The singlet state (denoted by subscript 0, and characterized by antiparallel spins of the proton and electron) and the triplet state (denoted by subscript 1, and characterized by parallel spins of the proton and electron), are separated by an energy difference of $5.9 \times 10^{-6}$ eV, corresponding to the temperature $T_{10} \approx 0.068$ K (and a wavelength of 21 cm). In equilibrium, the relative populations of these two states follow:

\begin{equation}
 \frac{n_1}{n_0} = \left(\frac{g_1}{g_0}\right) \exp\left[-\frac{T_{10}}{T_s}\right]
\end{equation} 
where the $g_1/g_0 = 3$ and $T_s$ is known as the spin temperature and characterizes the thermal equilibrium between the two states. 

 When the spin temperature is far greater than $T_{10} = 68$ mK (which corresponds to the energy difference between the two levels), we can approximate the previous equation as:
 \begin{equation}
n_1 \approx 3 n_0 \approx \frac{3 n_{\rm HI}}{4}        
\end{equation} 
where $n_{\rm HI}$ is the total number density of neutral hydrogen.

The intergalactic 21-cm line is typically observed in emission or absorption against the CMB, which can excite the hydrogen atoms from the singlet to the triplet state. The spin temperature of the neutral gas also changes due to collisional excitation and de-excitation processes and/or  Lyman-$\alpha$ coupling (the Wouthuysen-Field process, \cite{wouthuysen1952, field1958}). The expression for the spin temperature is given by:
\begin{equation}
 T_s^{-1} = \frac{T_{\rm CMB}^{-1} + x_c T_{K}^{-1} + x_{\alpha} T_{\alpha}^{-1}}{1 + x_c + x_{\alpha}}
\end{equation} 
where $T_K$ is the kinetic temperature of the gas, $T_{\rm CMB}$ is the CMB temperature, and $T_{\alpha}$ is known as the effective  Lyman-$\alpha$ color temperature, defined through the relation:
\begin{equation}
 \frac{P_{01}}{P_{10}} = 3 \left(1 - \frac{T_{01}}{T_c}\right)
\end{equation} 
where $P_{01}$ and $P_{10}$ are the spin excitation and de-excitation rates, respectively, from Lyman-$\alpha$ absorptions. In the high redshift IGM,  $T_c \to T_K$ due to the large number of Lyman-$\alpha$ scatterings. The coefficients $x_c$ and $x_{\alpha}$ are the coupling factors of the collisional and Wouthuysen-Field processes, respectively.

At very early epochs, the spin temperature tends to be in equilibrium with the ambient CMB temperature. When the gas density is high, collisions couple the spin temperature to the kinetic temperature. During the dark ages and reionization, the interplay of different processes such as heating of the gas, coupling of the gas temperature and the spin temperature, and the evolving ionization of the gas change the frequency structure of the signal (for a detailed review of the various processes and their impact on the signal, see Ref. \cite{furlanettorev}). This makes the spin temperature a powerful probe of the onset and timing of astrophysical processes during reionization.  
 
 The brightness temperature $T_b(\nu)$ of a radio source (assumed to be a neutral hydrogen cloud located at redshift $z$) in the Rayleigh-Jeans limit (an excellent approximation at these frequencies) can be expressed in terms of its intensity $I_{\nu}$ as:
 \begin{equation}
  T_b' = \frac{I_{\nu} c^2}{2 k_B \nu^2}
  \label{intensitytemperature}
 \end{equation} 
with $k_B$ being Boltzmann's constant. Using the equation of radiative transfer through a neutral hydrogen cloud having the spin temperature $T_s$, the brightness temperature of the cloud
is given by:
\begin{equation}
 T_b'(z) = T_{\rm CMB}(z) \exp(-\tau_{10}) + T_s (1 - \exp(-\tau_{10}))
\end{equation} 
 where $\tau_{10}$ is the optical depth for absorption of CMB photons and subsequent excitation from the singlet to the triplet level. It can be shown \cite{loeb2013, furlanettorev} that the optical depth is related to the number density of neutral hydrogen atoms,  $n_{\rm HI}$ by the expression:
 \begin{equation}
 \tau_{10} = \frac{3}{32 \pi} \frac{h c^2 A_{10}}{k_B T_s (z) \nu_{10}^2}\frac{n_{\rm HI} (z)}{(1+z) (dv_{\parallel}/dr_{\parallel})}
\end{equation} 
where $\nu_{10}$ is the frequency of the transition (1420 MHz) and $h$ is Planck's constant, and $dv_{\parallel}/dr_{\parallel}$ is the gradient of the proper velocity along the line of sight. 
From this, it can be shown \cite{loeb2013, furlanettorev} that the observed differential brightness temperature from a cloud at redshift $z$  against the CMB, given by $T_b(\nu) \equiv T_b'(\nu(1+z))/(1+z)$ (in the limit $\tau_{10} \ll 1$, $z \gg 1$) is given by:
\begin{equation}
 T_b(z)  = \frac{T_s - T_{\rm CMB}}{(1+z)} \tau_{10} \approx 28 \ {\rm{mK}} \left(\frac{\Omega_b h}{0.03}\right) \left(\frac{\Omega_m}{0.3}\right)^{-1/2} \left(\frac{1+z}{10}\right)^{1/2} x_{\rm HI} \left(1 - \frac{T_{\rm CMB}}{T_s}\right)
\end{equation} 
where $x_{\rm HI}$ is the neutral fraction of hydrogen, and the effects of peculiar velocities are neglected.

The above equation is suitable to describe the spin temperature evolution in the Cosmic Dawn and pre-reionization IGM ($z > 10$). From the above expression, we can see that that there is no signal if the spin temperature equals the CMB temperature. If $T_{\rm s} < T_{\rm CMB}$, the signal appears in absorption, and it appears in emission if $T_{\rm s} > T_{\rm CMB}$. A measurement of the brightness temperature was recently reported by the EDGES experiment \cite{bowman2018}) at $z \sim 17$. The Hydrogen Epoch of Reionization Array (HERA) recently \cite{hera2021} constrained the spin temperature of the $z \sim 8$ neutral IGM, placing limits on several physical models of reionization. It was also \cite{garsden2021} shown that the  Owens Valley Long Wavelength Array (OVRO-LWA) has sufficient sensitivity for a 21-cm detection at almost the edge of the Cosmic Dawn window, $z \sim 30$. 

In the post-reionization universe, the absorption is usually neglected as $T_s \gg T_{\rm CMB}$, and the neutral hydrogen fraction, $x_{\rm HI} \ll 1$ is usually  expressed as $x_{\rm HI} = \Omega_{\rm HI} (1+ \delta_{\rm HI})$ 
in terms of the comoving density parameter relative to the present-day critical density, $\Omega_{\rm HI} (z)$. This requires the substitution $x_{\rm HI}(z)  = \Omega_{\rm HI} (z) (1 + \delta_{\rm HI})$ where  $\delta_{\rm HI}$ is the overdensity of \HI. This gives \cite{bull2014, battye2012}:
 \begin{equation}
  T_b(z) = \frac{3 h c^3 A_{10} (1 + z)^2}{32 \pi k_B H(z) \nu_{10}^2}\frac{\Omega_{\rm HI} (z) \rho_{c,0} (1 + \delta_{\rm HI})}{m_H}
  \label{brtemp}
 \end{equation} 
 where where $\rho_{c,0} \equiv 3 H_0^2/8 \pi G$ is the present-day critical density.
\eq{brtemp} allows for the separation of the brightness temperature into a mean and a fluctuating component:
 \begin{equation}
   T_b(z) = \bar{T_b}(z) (1 + \delta_{\rm HI})
 \end{equation} 
and we have:
 \begin{equation}
  \bar{T_b}(z) = 44 \ \mu {\rm K} \left(\frac{\Omega_{\rm HI}(z) h}{2.45 \times 10^{-4}}\right) \frac{(1+z)^2}{E(z)}
  \label{tbar}
 \end{equation} 
 where $E(z) = H(z)/H_0$.
The difference in units between \eq{tbar} and the previous \eq{brtemp} ($\mu$K and mK) effectively reflects the difference in ambient neutral hydrogen densities prior to ($z >10$) and after ($z < 6$) reionization. In the post-reionization regime, it is relevant to study 
 the intensity \textit{fluctuations} of the brightness temperature, given by $\delta T_{\rm HI} = T_b(z) -  \bar{T_b}(z)$.
 In this regime, similar to the dark matter power spectrum, we can thus define the three-dimensional power spectrum of this intensity fluctuation, $P_{\rm HI} \equiv [\delta T_{\rm HI}(k,z)]^2$. We discuss the detailed method to evaluate this starting from the baryon population in haloes in the next section.
 
 \begin{figure}
\vskip-0.2in
 \includegraphics[width = \textwidth]{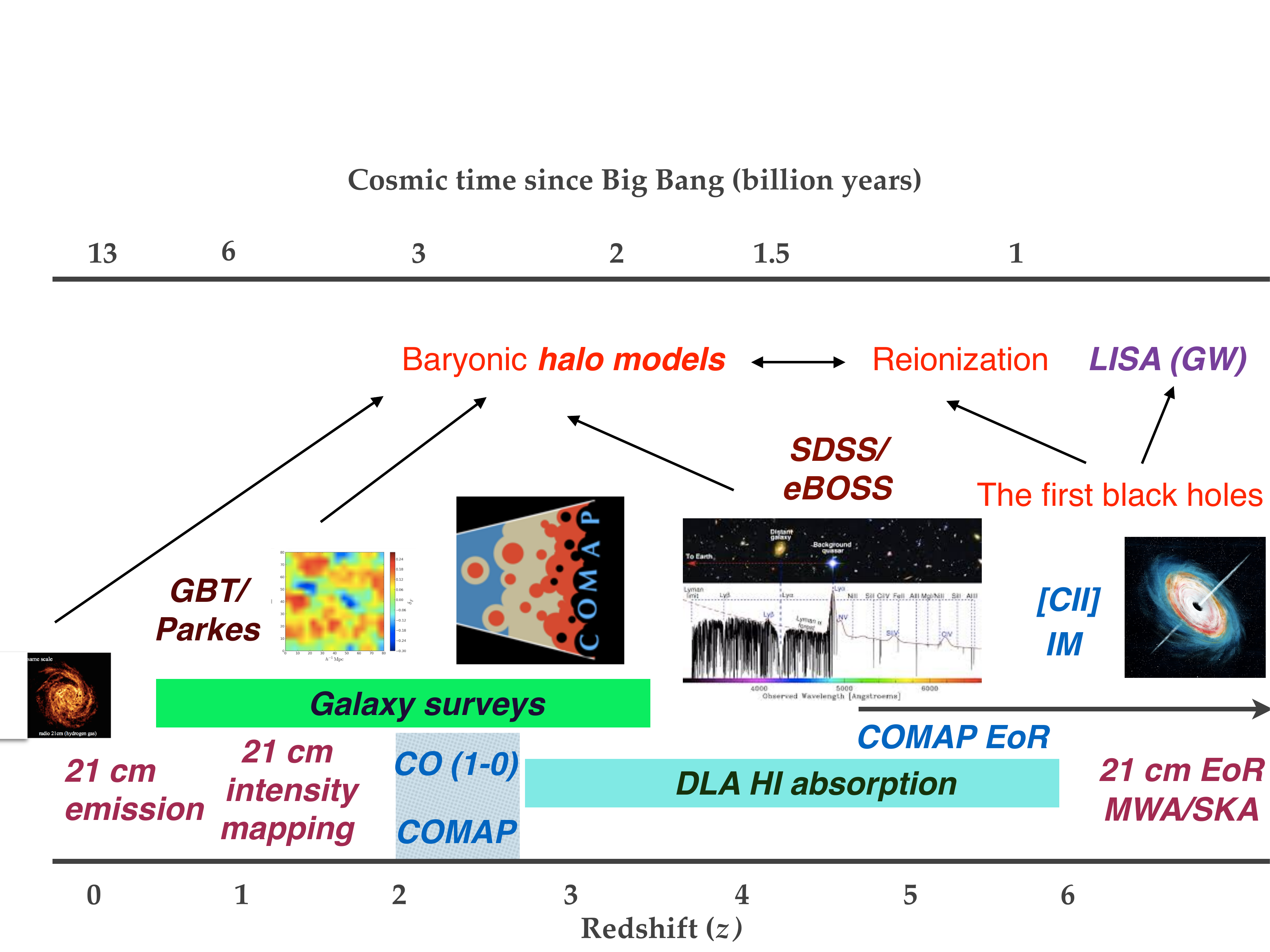}
\caption{Summary of the redshift ranges constrained by the different types of observations.}
\label{fig:summary}
\end{figure}

\subsection{Molecular gas and the sub-millimetre transitions}
\label{sec:submmintro}
In addition to the 21 cm transition of hydrogen considered above, there are several probes of the reionization and post-reionization epochs using molecular lines, such as those from the carbon monoxide (CO) molecule. CO, being an indicator of molecular hydrogen ($H_2$)  primarily traces star-forming regions \cite{breysse2014, mashian2015}, has a ladder of states covering rest frequencies of approximately $115 (J+1)$ GHz corresponding to its rotational transitions, where $J = 0, 1, ...$ is the quantum number of the transition. CO has been studied extensively from observations of local galaxies and recently also at intermediate redshifts, $z \sim 1-3$ \cite{keating2016, walter2016} and quasars out to as high as $z \sim 7$ \cite{venemans2017}. Accessing CO at the epoch of reionization is expected to offer valuable clues to the star formation and baryon cycle in the earliest galaxies. 

Two other tracers of interest in this regime are the  fine-structure line of singly ionized carbon (denoted by [CII]) with the rest wavelength 158$\mu$m (corresponding to 1.9 THz), which originates from a hyperfine transition between the $^{2}P_{3/2}$ and $^{2}P_{1/2}$
states of singly ionized carbon, and [OIII] line at $88 \mu$m, from the transition between the $^{3}P_{1}$ and $^{3}P_{0}$ states of doubly ionized oxygen.  The CII ion is a dominant coolant for the
interstellar medium, and the [CII] 158 $\mu$m line is the brightest far-infrared line and relatively easy to observe from the ground. The [OIII] line is commonly found in the vicinity of hot O-type stars,  and is senstitive to several physical properties of the ionized medium.  Both [CII] and [OIII] have been detected in galaxies out to $z \sim 9$ using the ALMA telescope \cite{pentericci2016, smit2018, laporte2017, harikane2020, tamura2019, hashimoto2018c, laporte2019, carniani2017}, and efforts are underway to observe them at these epochs without resolving individual galaxies \cite{terry2019, lagache2018, exclaimpaper2020}, the framework for which we describe in Sec. \ref{s:submmhalomodel}.

A schematic summary of all the tracers we consider, and their probes over various redshift ranges (which we discuss in detail in the forthcoming sections) is provided in Fig. \ref{fig:summary}.

\section{The halo model framework: from dark matter to baryons}\label{s:HIhalomodel}

As we have gathered, mapping the evolution of baryonic material across cosmic time using their line transitions promises deep insights into galaxy evolution, as well as theories of gravity and fundamental physics. The 21-cm line of neutral hydrogen, which is currently a useful probe of the \HI\ content of galaxies, is ideally suited for the bulk of the material, since hydrogen is the most abundant element in the Universe. 

Traditionally, galaxy surveys have been used as probes of the neutral hydrogen distribution \cite{zwaan05, zwaan2005a, martin12} at late times. The limits of current radio facilities, however, hamper the detection of 21-cm in emission from normal galaxies at early times. At these epochs, the HI distribution has been studied via the identification of Damped Lyman Alpha systems (DLAs, e.g., \cite{noterdaeme12, zafar2013, crighton2015}) — which are known to be the primary reservoirs of neutral hydrogen. 

A relatively new technique used to study HI evolution is known as intensity mapping. In this technique, the large-scale distribution of a tracer (like HI) can be mapped without individually resolving the galaxies which host the tracer. Being thus faster and less expensive than traditional galaxy surveys, intensity mapping of the 21 cm line had already been shown over the last two decades, e.g., \cite{chang10, wyithe2008a, anderson2018} to have the potential to provide constraints on cosmological parameters which are competitive with those from next generation experiments. 

Given the deluge of data expected from forthcoming radio observations, especially with the intensity mapping technique, it is therefore
timely and important to unify the data into a self-consistent framework that addresses both high- and low-redshift observations. Such a framework, which encapsulates the astrophysical information contained in HI (and, in general, baryonic gas tracers) into a set of key physical parameters, is crucial to properly take into account astrophysical uncertainties in order to place the most realistic constraints on Fundamental Physics and cosmology. Furthermore, it enables the most effective comparison to physical models of galaxy astrophysics, allowing us to unravel several outstanding questions in their evolution.

In a map of unresolved line emission over a large area in the sky, the main quantity of interest is the three-dimensional power spectrum, introduced for \HI\ in Sec. \ref{sec:21cmintro} as $P_{\rm HI} (k,z)$. This power spectrum is a measure of the strength of the intensity fluctuations as a function of wave number ($k$) at redshift $z$. To compute the power spectrum, one needs to establish a framework connecting dark matter and baryons. The most natural method to achieve this is via a data-driven halo model framework \cite{hpar2017}  for the evolution of baryons as tracers of large scale structure. This can be achieved by combining the different datasets available (e.g. \cite{hptrcar2015}), and bringing together the models available in the literature \cite{hptrcar2016} into a comprehensive picture. In so doing, it is found that there are two main ingredients required to connect the baryonic quantities (mass, luminosity) to the underlying dark matter halo model (described in Sec.\ref{sec:introcosmo}), namely:
(i) the average mass (or luminosity) associated with a dark matter halo of mass $M$ at redshift $z$, and (ii) if relevant, the distribution of the mass as a function of scale, $r$ from the centre of the dark matter halo. The latter quantity is relevant when the data constrains the detailed small-scale structure of the baryons, as for the case of \HI. We will now  describe this framework, closely paralleling the development of the dark matter halo model framework in Sec. \ref{sec:introcosmo}.

Given a prescription for populating the halos with baryons, e.g., $M_{\rm HI}(M)$ (for the case of \HI), defined as the average mass of \HI\ contained in a halo of mass $M$, we begin by computing the neutral hydrogen density, $\bar{\rho}_{\rm HI}(z)$ in terms of the dark matter mass function $n(M,z)$ as:
\begin{equation}
\bar{\rho}_{\rm HI}(z)= \int_{M_{\rm min}}^{\infty} dM n(M,z) M_{\rm HI}(M) \,
\end{equation}
in which $M_{\rm min}$ is an astrophysical parameter that quantifies the minimum mass of the dark matter halo that is able to host \HI. In practice, this minimum mass requirement is usually built-in to the $M_{\rm HI}(M)$ function itself, by using a suitable cutoff scale.
For quantifying large-scale clustering, we define the large-scale bias parameter $b_{\rm HI} (z)$, that describes how strongly \HI\ is clustered relative to the dark matter,  as:
\begin{equation}
b_{\rm HI} (z) = \frac{1}{\bar{\rho}_{\rm HI}(z)}\int_{M_{\rm min}}^{\infty} dM
n(M,z) b(M) M_{\rm HI}(M).
\label{biasHI}
\end{equation}
To describe small-scale clustering, we define the normalized Hankel transform of the density profile $\rho_{\rm HI} (r)$:
 \begin{equation}
 u_{\rm HI}(k|M) = \frac{4 \pi}{M_{\rm HI} (M)} \int_0^{R_v} \rho_{\rm HI}(r) \frac{\sin kr}{kr} r^2 \ dr
\end{equation}
analogously to \eq{udm} for dark matter. The 
normalization of $u(k|M)$ above is to the total \HI{} mass, i.e. $M_{\rm HI} (M)$. In most of the standard analyses, the baryonic profile is assumed truncated at the halo virial radius, which is taken to be a standard `scale' cutting off the halo. The corresponding scale radius for the \HI\ is defined through $r_s = R_v(M)/c_{\rm HI}$ where $c_{\rm HI}$ is known as the  concentration parameter, analogous to the corresponding one for dark matter defined in \eq{concparamdm}. \footnote{Profiles like the NFW diverge as $\rho(r) \sim r^{-3}$ as $r \to \infty$, causing a logarithmic divergence of the mass contained within them. Thus, it is helpful to truncate the profile at the virial radius which defines the `boundary' enclosing the total mass, just as done for the dark matter case. Recent studies advocate the use of the \textit{splashback} radius, at which the accreted material reaches its first apocenter after turnaround \cite{diemer2018} and is separated from the virial radius by factors of a few, as a more precise definition of the halo boundary.}

The dark matter framework is then followed to compute the one- and two halo terms of the \HI{} power spectrum, given by:
\begin{equation}
P_{\rm 1h, HI} =  \frac{1}{\bar{\rho}_{\rm HI}^2} \int dM \  n(M) \ M_{\rm HI}^2 \ |u_{\rm HI} (k|M)|^2
\label{onehalo}
\end{equation}
and
\begin{equation}
P_{\rm 2h, HI} =  P_{\rm lin} (k) \left[\frac{1}{\bar{\rho}_{\rm HI}} \int dM \  n(M) \ M_{\rm HI} (M) \ b (M) \ |u_{\rm HI} (k|M)| \right]^2
\label{twohalo}
\end{equation}
which are analogous to \eq{powerspecdm}.
Some studies in the literature \cite{wolz2019, villaescusa2018} additionally model the shot noise of the halo power spectrum, $P_{\rm SN}$, which may be computed as:

\begin{equation}
   P_{\rm SN} =  \frac{1}{\bar{\rho}_{\rm HI}^2} \int dM \  n(M) \ M_{\rm HI}^2 \ 
   \label{shotnoiseHI}
\end{equation}
and corresponds to the $k \to 0$ limit of the one-halo power spectrum. This term provides a rough measure of the  Poissonian noise due to the finite number of discrete sources (such as HI-bearing galaxies).
In practice, however, in the redshift range considered in the context of \HI\ intensity
mapping, the shot noise contribution is expected to be negligible, e.g., \cite{seo2010, seehars2016} and relatively unconstrained by observations, so it is not relevant in a data-driven approach. It will, however, become important in the context of submillimetre observations described in the next section.

In all of the above, the dark matter mass function $n(M, z)$ can be computed from analytical arguments or simulations. In most of our analyses in the following sections, we will assume the Sheth - Tormen form defined in Sec. \ref{sec:astrocosmo}, though various other forms are possible \cite{tinker2008, shirasaki2021}.
 
Finally, the neutral hydrogen fraction, required for calculating the brightness temperature as described in Sec. \ref{sec:21cmintro}, is computed as:
\begin{equation}
 \Omega_{\rm HI} (z) = \frac{\bar{\rho}_{\rm HI}(z)}{\rho_{c,0}}
 \label{omegaHIanalyt}
\end{equation} 
where $\rho_{c,0} \equiv 3 H_0^2/8 \pi G$ is the critical density of the Universe at redshift 0.

In several recent observations, the real-space correlation function is measured, which is defined as the Fourier transform of the power spectrum ($P_{\rm HI, 1h} + P_{\rm HI, 2h})$. It is computed as:
\begin{equation}
\xi_{\rm HI} (r) = \frac{1}{2 \pi ^2} \int k^3 (P_{\rm 1h, HI} + P_{\rm 2h, HI}) \frac{\sin kr}{kr} \frac{dk}{k}
\end{equation}

The above discussion completes the generic treatment for analysing observations of baryonic gas in a halo model framework, given the tracer-to-halo mass relation and its small-scale profile. Note that the above framework is most naturally suited to describe intensity mapping observations, since it is a mass- (or luminosity) weighted measurement analogous to the approach for dark matter, similar to that employed in studies of the cosmic infrared background, e.g., \cite{fernandez2006}. This is different from, e.g., a number-counts weighted approach which is commonly used in surveys of discrete tracers like galaxies, e.g., \cite{cooraysheth2002} with a halo occupation distribution (HOD). While number-count- and mass-count-weighted frameworks are completely equivalent in their capture of small-scale effects (via the `satellite' galaxies distribution and the 1-halo term respectively), it is more natural to describe intensity mapping observations using a mass-weighted approach, since we are dealing here with diffuse gas not all of which may be located inside galaxies and as such does not form a discrete set. This also facilitates the extension of this parameterization to the reionization regime and beyond. Quantifying the gas content in discrete systems (such as galaxies and Damped Lyman-Alpha systems) is made possible due to the flexibility in mass and cutoff scales in the framework, as we show below by describing how specific expressions that arise in the context of \HI\ observations from galaxy surveys and DLAs can be modelled with the above ingredients.

\subsection{21 cm from galaxy surveys}
\label{sec:21cmgal}

The main observable in a survey of galaxies detected in 21 cm is their \textit{mass function}, which measures the volume density of \HI-selected galaxies as a function of the \HI\ mass in different mass bins. Denoted by $\phi(M_{\rm HI})$, it is typically found to follow a functional form of the type \cite{martin10, zwaan05}:
\begin{equation}
 \phi(M_{\rm HI}) \equiv \frac{dn}{d \log_{10} M_{\rm HI}} =  \frac{\phi_*}{M_0} \ \left(\frac{M_{\rm HI}}{M_0} \right)^{-a} e^{-M_{\rm HI}/M_0}
\end{equation} 
where $\phi_*, a$ and $M_0$ are free parameters.
Given the HI mass - halo mass relation $M_{\rm HI}(M,z)$, the above mass function can be modelled by:
\begin{equation}
    \phi(M_{\rm HI}) = \frac{dn}{d \log_{10} M} \left|\frac{d \log_{10} M}{d \log_{10} M_{\rm HI}}\right|
    \label{phifrommhi}
\end{equation}
with the first term calculated from $n(M,z)$.

The \HI\ mass function is then used to calculate the mass density of \HI\ in galaxies:
\begin{equation}
 \rho_{\rm HI} = \int M_{\rm HI} \phi(M_{\rm HI}) d M_{\rm HI} = M_0 \phi_* \Gamma(2 - a) 
\end{equation}
from which the density parameter of \HI\ in galaxies, $\Omega_{\rm HI, gal} = $ can be calculated:
\begin{equation}
  \Omega_{\rm HI, gal} =  \rho_{\rm HI}/\rho_{c,0}
\end{equation}

The clustering of HI-selected galaxies can generally be computed following \eq{biasHI}, with the modification that the lower limits in the integrals in \eq{biasHI} are fixed to the halo mass $M_{\rm min}$ corresponding to the minimum \HI\ mass observable by the survey. Thus, the \HI\ bias measured from galaxy surveys is modelled by:
\begin{equation}
b_{\rm HI,gal} (z) = \frac{\int_{M_{\rm min}}^{\infty} dM \ n(M,z)\ b (M,z) \ M_{\rm HI} (M,z)}{\int_{M_{\rm min}}^{\infty} dM \  n(M,z) \ M_{\rm HI} (M,z)}
\label{biasHIgal}
\end{equation}

We note in passing that, given the \HI\ mass function and reversing the procedure described in \eq{phifrommhi}, commonly known as the \textit{abundance matching} technique in the literature, we can directly estimate $M_{\rm HI}(M)$ from observational data. This makes the assumption that $M_{\rm HI}(M)$ is a monotonic function of $M$, and is given by solving the equation \cite{vale2004}:
\begin{equation}
    \int_{M(M_{\rm HI})}^{\infty} \frac{dn}{d \log_{10} M'} d \log_{10} M' = \int_{M_{\rm HI}}^{\infty} \phi(M_{\rm HI'}) d \log_{10} M_{\rm HI}'
    \label{abmatchhi}
\end{equation}
and an appropriate functional form is fitted to the resulting datapoints $\{M_{\rm HI}, M\}$. Such an approach, for the case of \HI\ and using the mass functions from recent literature \cite{martin10, zwaan05} is found to lead to results which are consistent \cite{hpgk2017} with the direct fitting to observations described below.

\subsection{Neutral hydrogen  from Damped Lyman Alpha systems}
\label{sec:dlahimodels}

As we have seen in Sec. \ref{sec:astrocosmo}, Damped Lyman Alpha systems (DLAs) represent the highest column density \HI-bearing systems in the IGM.
In a survey measuring the neutral hydrogen fraction using DLAs, the primary observable is called the \textit{column density distribution function} defined in \eq{coldens} denoted by $f_{\rm HI} (N_{\rm HI})$, where $N_{\rm HI}$ is the column density of the DLAs:
\begin{equation}
 d^2 \mathcal{N} = f_{\rm HI} (N_{\rm HI}, X) dN dX
\end{equation} 
with $\mathcal{N}$ being the observed incidence rate of DLAs in the absorption interval $dX$ and the column density range $dN_{\rm HI}$. 
To model the column density distribution using the framework described in the previous section, the hydrogen density profile $\rho_{\rm HI}(r)$ as a function of $r$ is first used to calculate the column density of DLAs using the relation:
\begin{equation}
 N_{\rm HI}(s) = \frac{2}{m_H} \int_0^{\sqrt{R_v(M)^2 - s^2}} \rho_{\rm HI} (r = \sqrt{s^2 + l^2}) \ dl 
 \label{coldenss}
\end{equation} 
where $s$ is the impact parameter of a line-of-sight through the DLA. 

The cross-section for a system to be identified as a DLA, $\sigma_{\rm DLA}$ can then be computed by 
\begin{equation}
    \sigma_{\rm DLA} = \pi s_*^2
\end{equation}
where $s_*$ is the root\footnote{If no positive root $s^*$ exists, it physically means that the column density in the line-of-sight does not reach $N_{\rm HI} = 10^{20.3}$ cm$^{-2}$ even at zero impact parameter, so the cross-section is zero for such systems. Hence, in such cases, $s ^{*}$ is set to zero.}  of the equation $N_{\rm HI}(s_*) = 10^{20.3}$ cm$^{-2}$ (which is the column density threshold for the appearance of DLAs). 

Given the DLA cross-section, 
the column density distribution $f_{\rm HI}(N_{\rm HI}, z)$ is modelled by:
\begin{equation}
 f(N_{\rm HI}, z) = \frac{c}{H_0} \int_0^{\infty} n(M,z) \left|\frac{d \sigma}{d N_{\rm HI}} (M,z) \right| \ dM 
 \label{coldensdef}
\end{equation} 
where the $d \sigma/d N_{\rm HI} =  2 \pi \ s \ ds/d N_{\rm HI}$, with $N_{\rm HI} (s)$ defined as in \eq{coldenss}.

The clustering of DLAs is captured by the DLA bias, $b_{\rm DLA}$  defined by:
\begin{equation}
 b_{\rm DLA} (z) =  \frac{\int_{0}^{\infty} dM n (M,z) b(M,z) \sigma_{\rm DLA} (M,z)}{\int_{0}^{\infty} dM n (M,z) \sigma_{\rm DLA} (M,z)}.
 \label{bdla}
\end{equation} 
Note that the above expression is almost identical to \eq{biasHI}, with the only difference being the weighting of the bias by the cross-section of DLA absorbers.

Another observable is the incidence of the DLAs, denoted by $dN/dX$ which quantifies the number of systems per absorption path length, and is calculated as:
\begin{equation}
 \frac{dN}{dX} = \frac{c}{H_0} \int_0^{\infty} n(M,z) \sigma_{\rm DLA}(M,z) \ dM
 \label{dndxdef}
\end{equation} 
Finally, from the column density distribution, the density parameter of hydrogen in DLAs can be calculated as:
\begin{equation}
 \Omega_{\rm HI}^{\rm DLA} = \frac{m_H H_0}{c \rho_{c,0}} \int_{N_{\rm HI, min}}^{\infty} N_{\rm HI} f_{\rm HI} (N_{\rm HI}, X) dN_{\rm HI} dX \, ,
\end{equation} 
in which the lower limit\footnote{The lower limit changes to $10^{19}$ cm$^{-2}$, \cite{zafar2013} in case the sub-DLAs too are accounted for while calculating the gas density parameter. Lower column-density systems, such as the Lyman-$\alpha$ forest,  make negligible contributions to the total gas density.} of the integral is set by the column density threshold for DLAs, i.e. $N_{\rm HI, min} = 10^{20.3}$ cm$^{-2}$.

In alternate approaches to modelling DLAs \cite{villaescusa2018}, the cross section $\sigma_{\rm DLA} (M)$ itself may directly be modelled using a functional form, and the DLA quantities calculated from $\sigma_{\rm DLA}$. Various simulation-based approaches have also been used to quantify the neutral hydrogen at different redshifts from DLAs \cite{pontzen2008, bird2014}.

\subsection{HI-halo mass relations and density profiles}
\label{sec:analytical}
We have seen in the above discussion that the two key inputs needed in the halo model framework for hydrogen are $M_{\rm HI}(M)$, the prescription for
assigning \HI\ to the dark matter haloes, and $\rho_{\rm HI}(r)$, which describes how this mass is distributed as a function of scale.  

Several forms had been used to model the $M_{\rm HI}(M)$ function in the past literature. At redshifts probed chiefly by DLA observations, $z \sim 2-5$, a relation between HI mass and halo mass of the form 
$M_{\rm HI} = \alpha M \exp(-M/M_0)$
where $\alpha$ is a constant of normalization, and $M_0$ is a lower mass cutoff was commonly used\cite{barnes2014}. 
At lower redshifts, various forms had been proposed, among which are $M_{\rm HI} = f M$ which is a constant fraction of the halo mass between a fixed lower limit in halo mass, $M_{\rm min}$ and upper limit $M_{\rm max}$ \cite{bagla2010}. 

Reconciling the high- and low-redshift approaches \cite{hptrcar2016} leads to some fascinating insights about the occupation of \HI\ in dark matter haloes. Specifically, it is found that in order to match all the current observational data (DLAs, 21 cm galaxy surveys and intensity mapping observations) over $z \sim 0-5$, the connection between HI mass and halo mass needs to be modelled using a function of the form:
\begin{eqnarray}
M_{\rm HI} (M) &=& \alpha f_{\rm H,c} M \left(\frac{M}{10^{11} h^{-1} M_{\odot}}\right)^{\beta} \exp\left[-\left(\frac{v_{c0}}{v_c(M)}\right)^3\right] \nonumber \\
\end{eqnarray}
We now describe the form of this expression in some detail. 
It involves three free parameters, $\alpha$, $\beta$ and $v_{c,0}$:

(i) $\alpha$ is the  overall normalization of the HI-halo mass relation. Physically, it represents the fraction of HI, relative to the cosmic fraction ($f_{\rm H,c}$) that resides in a halo of mass $M$ at redshift $z$. The cosmic fraction is the primordial hydrogen fraction by mass, defined as:
\begin{equation}
f_{\rm H,c} = (1 - Y_{\rm He}) \Omega_b/\Omega_m \, ,
\label{fhc}
\end{equation}
in which $Y_{\rm He} = 0.24$ is the primordial helium fraction.

(ii) $\beta$ is the logarithmic slope of \HI\ mass to halo mass. Any nonzero value of $\beta$ describes a  departure from proportionality of the HI mass and halo mass. The $\beta$ is physically connected to physical processes that deplete \HI\ in galaxies (such as quenching and feedback from the intergalactic medium).

(iii) The parameter $v_{\rm c, 0} (M)$  represents a lower cutoff to the HI-halo mass relation. It describes the minimum mass (or equivalent circular velocity) of a halo able to host neutral hydrogen. In the above equation, the circular velocity is calculated through:
\begin{equation}
    v_{c} = \sqrt\frac{GM}{R_v(M)}
    \label{vcRv}
\end{equation}
where $R_v(M)$ is the virial radius defined in \eq{virialradius}. 
\footnote{An equivalent representation of the lower cutoff is to use a minimum halo mass ($M_{\rm min} (z)$) in place of the circular velocity. Since we see from \eq{vcRv} and \eq{virialradius} that $v_c \propto M^{1/3}(1+z)^{1/2}$, the exponential term  can be written as  $\exp(-M/M_{\rm min} (z))$ if we are using a mass cutoff. However, using the circular velocity -- instead of halo mass  -- to denote the cutoff makes the physical connection to the intergalactic ionizing background more natural, as we will see in the next section.}

Just like the HI-halo mass relation, it is also important to parametrize the profile of the neutral hydrogen in the dark matter halo, $\rho_{\rm HI}(r)$. In the literature, this function had been modelled by an altered version of the NFW profile introduced in Sec. \ref{sec:astrocosmo} \cite{maller2004, barnes2014, hpar2017}:
\begin{equation}
\rho_{\rm HI} (r) = \frac{\rho_0 r_{\rm s, HI}^3}{(r + 0.75 r_{\rm s, HI}) (r+r_{\rm s, HI})^2}
\label{rhodefnfw}
\end{equation}
The quantity $r_{\rm s,HI}$ is the scale radius of \HI, which is defined as $r_{\rm s, HI} \equiv R_v(M)/c_{\rm HI}$, introducing a concentration parameter $c_{\rm HI}$ analogous to the one for dark matter in \eq{concparamdm}:
\begin{equation}
    c_{\rm HI}(M,z) =  c_{\rm HI, 0} \left(\frac{M}{10^{11} M_{\odot}} \right)^{-0.109} \frac{4}{(1+z)^{\gamma}}
    \label{concparamhi}
\end{equation}
We thus see that the profile function brings two additional parameters to the model: $c_{\rm HI, 0}$ and $\gamma$. Of these, $c_{\rm HI}$ is the overall normalization of the concentration, and $\gamma$ describes how it evolves with redshift. \footnote{Note that the profile has a negligible halo mass dependence (the power of -0.109). This is inherited from the dark matter framework\cite{duffy}  and does not have any significant bearing on the modelling.}
Recent data, especially from HI-rich disk galaxies \cite{bigiel2012} at low redshifts, favours the profile for \HI\ in haloes being of the exponential form (rather than the modified NFW relation above):
\begin{equation}
    \rho(r,M) = \rho_0 \exp(-r/r_{\rm s, HI})
\label{rhodefexp}
\end{equation}
In both forms of the profile, the parameter $\rho_0$ is fixed by normalization to the total HI mass, $M_{\rm HI}(M)$, just as done for  the case of the dark matter halo model.
The profile and the HI-halo mass relation can now be used to compute the 1- and 2-halo terms of the power spectrum defined in \eq{onehalo} and \eq{twohalo} respectively. For this, it is necessary to compute the normalized Hankel transform of the profile function, which can be expressed analytically for both the profile choices above \cite{hparaa2017}.
As an explicit example, for  the \HI\ profile in the exponential form, the  expression for the normalized Hankel transform is given by:
\begin{equation}
u_{\rm HI}(k|M) = \frac{4 \pi \rho_0 r_{\rm s, HI}^3 u_1(k|M)}{M_{\rm HI} (M)}
\end{equation}
where
\begin{equation}
u_1(k|M) = \frac{2}{(1 + k^2 r_{\rm s, HI}^2)^2},
\end{equation}
which can then be used to compute the full power spectrum.

The five parameters  $\{c_{\rm HI, 0}, \alpha, \beta, v_{c, 0}, \gamma\}$ can now be fixed by fitting the framework above to the current astrophysical data describing \HI.  This is achieved using a Markov Chain Monte Carlo (MCMC) approach \cite{hparaa2017}  and the resulting best-fitting parameters and their uncertainties are summarized in the first column of Table \ref{table:constraints}. There are a few salient points that emerge from the analysis:
\begin{enumerate}
    \item The best fitting value of $\alpha$ is $\alpha = 0.09$. This denotes that about $10\%$ of the hydrogen is in atomic form over the post-reionization Universe, and is in line with recent simulations and observations that predict the fraction of cold gas (i.e. the sum of the atomic and molecular components) to be in the range of $\sim 10-20$\% in low-redshift galaxies\cite{stern2016}. Interestingly, the observations do not favour an evolution in $\alpha$ with redshift. This in in line with the findings from DLA studies \cite{prochaska09}, which indicate evidence for non-evolution of hydrogen in galaxies over $z \sim 0-5$, and reiterates the role of HI as an `intermediary' in the baryon cycle \cite{wang2020, bouche2010, lilly2013, hploebsfr2020}: the HI replenishment from the IGM is compensated by its conversion to molecular hydrogen, $H_2$ which is used up by star formation.
    
    \item The slope, $\beta$ is found to have the (negative) value $\beta = -0.58$. It is found that the slope is dominantly influenced by the form of the HI mass function observed at low redshifts for which exquisite constraints are available \cite{zwaan05, martin10}. The suppression of the resultant HI-halo mass slope from unity is in line with evidence for quenching, or the suppression of star formation in massive haloes due to feedback from the IGM \cite{birnboim2007, finlator2017}.
    
    \item The cutoff $v_{\rm c,0}$ has the value 36.3 km/s. This can be directly related to the circular speed of suppression of dwarf galaxies by the ambient ionizing background in the IGM, which was worked out in the mid-80s and 90s from analytical arguments balancing ionization and recombination \cite{rees1986, efstathiou1992, quinn1996}. Fascinatingly, such an analysis favours a value of $v_c \sim 37$ km/s, almost perfectly matched to the value obtained by combining the latest ($\sim$ 2017)  \HI\ observations today! This greatly strengthens the case for using circular velocity as a lower cutoff for \HI\ in haloes, and sheds light on the physics associated with this parameter.

\end{enumerate}

\subsection{The sub-millimetre regime}
\label{s:submmhalomodel}

Although hydrogen is the most abundant element in the Universe, there are several exciting prospects for making intensity maps of other salient lines, an important example being molecular lines, like the carbon monoxide (CO) lines introduced in Sec. \ref{sec:submmintro}. CO is the second most abundant molecule in the universe (after molecular hydrogen) and much easier to detect from ground-based experiments. CO behaves as a tracer of molecular hydrogen, which has no permanent dipole moment due to symmetry and thus no rotational transitions of its own.  CO lines are thus the primary way to trace molecular gas within and outside our galaxy, and very sensitive to the spatial distribution of star formation \cite{hploebsfr2020}. The CO line corresponding to the transition between rotational quantum numbers $J$ and $J - 1$ has a rest frequency of approximately $J \times$115.27 GHz, making it an ideal target in the sub millimetre regime. It is easy to separate the signal from contaminants due to its multiple emission lines (with different values of the angular momentum quantum number $J$) which have a  well defined frequency relationship, a feature that is not available to other tracers. This also enables effective cross-correlation between observations at lower and higher frequency bands which are integral multiples of each other (e.g., a  frequency band covering 26-34 GHz will be sensitive to the CO 2-1 line at $z \sim$ 6-8, while also capturing the CO 1-0 transition at $z \sim$ 2-3.)

The CO Mapping Array Project (COMAP) aims to detect the CO molecule in emission during the epoch of Galaxy Assembly, about two billion years after the Big Bang. Science observations with the COMAP Pathfinder began in 2019 using a 19-pixel 26-34 GHz receiver mounted on a 10.4 metre dish at the  Owens Valley Radio Observatory.  It was shown \cite{li2015} using simulations, that the COMAP experiment is capable of providing close to 8$\sigma$ constraints on the CO intensity power spectrum at large scales. In November 2021, the COMAP Pathfinder released the first direct 3D measurement of the CO power spectrum on large scales \cite{cleary2021}, nearly an
              order of magnitude improvement compared to the previous
              best measurement \cite{keating2020}. 

Observations made using the Karl G. Jansky Very Large Array (JVLA) and the Atacama Large Millimeter Array (ALMA) have recently shown that line emission from the CO transitions will be bright even at high redshift, $z > 6$.  The levels of foreground contamination in a CO survey are also much lower than for many other types of line intensity mapping, making it a promising target for ground-based observations. Recently, the CO Power Spectrum Survey [COPSS; Ref. \cite{keating2016}] detected, for the first time, the aggregate CO intensity in emission from galaxies at the peak of the star formation history of the universe. The mmIME experiment recently \cite{keating2020} announced the detection of unresolved intensity from the aggregate CO (3-2) emission in galaxies at $z \sim 2.5$ in the shot noise regime. In \cite{pullen2018}, confirmed by \cite{yang2019}, there was a tentative detection of the 158 micron line of [CII], an excellent tracer of star formation, reported, for the first time, by combining Planck CMB maps with  quasars from the BOSS and CMASS galaxy surveys.

By developing a framework that can incorporate current observational constraints on the abundances and clustering of the tracers, we can readily use the wealth of upcoming submillimetre observations to constrain galaxy evolution. Similar to the case of \HI, the main observable in the case of submillimetre intensity mapping observations is the power spectrum. However, in contrast to the \HI\ case, the luminosity of the tracer is normally used instead of the mass, so the prescription to be modelled is, e.g.,  $L_{\rm CO} (M,z)$ in the case of CO. Another difference is that the modelling of sub-millimetre intensity mapping usually focuses on the linear regime alone (since the data typically does not constrain the behaviour of the profile of the tracer). Instead of modelling the  full one-halo term, what is typically modelled is the contribution from the \textit{shot noise} alone, which is analogous to the term introduced briefly in the previous section in \eq{shotnoiseHI}.

To compute the power spectrum, we use the specific intensity of a submillimetre line  observed at a frequency, $\nu_{\rm obs}$, given by:
\begin{equation}
 I(\nu_{\rm obs}) = \frac{c}{4 \pi} \int_0^{\infty} dz' \frac{\epsilon[\nu_{\rm obs} (1 + z')]}{H(z') (1 + z')^4}
\end{equation} 
in which $\epsilon[\nu_{\rm obs} (1 + z')]$ is known as the volume emissivity of the emitted line.
It is usually assumed that the profile of each line is a delta function \footnote{The effects of line broadening on the intensity and power spectra are analytically treated in Ref. \cite{chung2021lb}.} at the rest frequency $\nu_{\rm em}$. This implies that the emissivity can be expressed as an integral of the host halo mass $M$:
\begin{equation}
 \epsilon(\nu, z) = \delta_D(\nu - \nu_{\rm em}) (1 + z)^3 f_{\rm duty} \int_{M_{\rm min}}^{\infty} dM \frac{dn}{dM} L(M,z)
 \label{emissivity}
\end{equation} 
where $L (M,z)$ is the luminosity of the line under consideration, and it is assumed that a fraction $f_{\rm duty}$ (usually called the `duty-cycle' factor) of all haloes above a mass $M_{\rm min}$ contribute to the observed emission  in the line of interest, e.g., \cite{lidz2011}. This parameter can be approximated by $f_{\rm duty} = t_s/t_H$, where $t_s$ is the star formation timescale and $t_H$ is the Hubble time at the redshift under consideration. 
It is to be noted that both of these parameters are fairly poorly constrained by the data;  $f_{\rm duty}$ in particular is found to differ by more than an order of magnitude between different models \cite{pullen2013, keating2016}. For this reason, $f_{\rm duty}$ is alternatively taken into account by introducing intrinsic scatter parameters \cite{li2015} to account for the differences in the star formation activity of haloes.

Using \eq{emissivity}, the specific intensity can be rewritten as:
\begin{equation}
I(\nu_{\rm obs}) = \frac{c}{4 \pi} \frac{1}{\nu_{\rm em} H(z_{\rm em})}  f_{\rm duty} \int_{M_{\rm min}}^{\infty} dM \frac{dn}{dM} L(M,z)
\label{COspint}
\end{equation} 
where $z_{\rm em}$ is the redshift of the emitting source.
Just as in the case of \HI,
the brightness temperature, $T$  can be derived from the specific intensity through the relation $I(\nu_{\rm obs}) = 2 k_B \nu_{\rm obs}^2 T /c^2$. 
From this, the expression for the brightness temperature becomes:
\begin{equation}
T(z) = \frac{c^3}{8 \pi}\frac{(1 + z_{\rm em})^2}{k_B \nu_J^3 H(z_{\rm em})} f_{\rm duty} \int_{M_{\rm min}}^{\infty} dM \frac{dn}{dM} L(M,z)
\label{tco}
\end{equation} 
The clustering of the submillimetre sources  can be modelled in analogy with the \HI\ case by weighting the dark matter halo bias by the tracer luminosity-halo mass relation.
The expression for the clustering is therefore given by:
\begin{equation}
 b_{\rm submm}(z) = \frac{\int_{M_{\rm min}}^{\infty} dM (dn/dM) L (M,z) b(M,z)}{\int_{M_{\rm min}}^{\infty} dM (dn/dM) L (M,z)}
\end{equation} 
in which we see that $L(M,z)$ has now taken the place of $M_{\rm HI}(M,z)$ in \eq{biasHI}.

The shot noise contribution to the power is expressed as:
\begin{equation}
 P_{\rm shot}(z) = \frac{1}{f_{\rm duty}}\frac{\int_{M_{\rm min}}^{\infty} dM (dn/dM) L (M,z)^2}{\left(\int_{M_{\rm min}}^{\infty} dM (dn/dM) L (M,z)\right)^2}
\end{equation} 
The total power spectrum of the intensity fluctuations is the sum of the clustering (two-halo) and shot-noise components:
\begin{equation}
 P_{\rm submm}(k,z) =  T (z)^2 [b_{\rm submm}(z)^2 P_{\rm lin}(k,z) + P_{\rm shot}(z)]
 \label{submmpower}
\end{equation} 
in which $P_{\rm lin}(z)$ denotes the dark matter power spectrum calculated in linear theory. \footnote{\eq{submmpower} assumes that the power is measured in units of ${\rm K}^2$, alternatively, it can directly be measured in units of Jy/sr$^2$ (as commonly done for the cases of [CII] and [OIII]) in which case the intensity in \eq{COspint} is  used: $P_{\rm submm}(k,z) = I(\nu_{\rm obs})^2 b_{\rm submm}^2(z) P_{\rm lin}(k,z) + P_{\rm shot}(z)$.}
Several times,  we need to plot the power spectrum in logarithmic $k$-bins, which is given by:
\begin{equation}
 \Delta_{k}^2(z) = \frac{k^3  P_{\rm HI/submm}(k,z)^2}{2 \pi^2}
 \label{COpowspeclog}
\end{equation} 
 which is equally valid for both the HI power spectrum in \eq{onehalo} and \eq{twohalo}, and the submillimetre one in \eq{submmpower}.
 
To model the main observable for submillimetre intensity mapping, the abundance matching technique (introduced at the end of Sec. \ref{sec:21cmgal}) is found to be suitable for both CO and [CII] at $z \sim 0$, due to the availability of the luminosity functions at low redshifts for both these tracers \cite{hemmati2017, keres2003}, which is denoted by $\phi(L)$ and measures the number density of CO- or CII-luminous galaxies in logarithmic luminosity bins.\footnote{The data do not constrain a non-monotonic behaviour of the luminosity-halo mass relations, so this is a reasonable approach.} Specifically, \eq{abmatchhi} gets modified to:
\begin{equation}
    \int_{M(L)}^{\infty} \frac{dn}{d \log_{10} M'} d \log_{10} M' = \int_{L}^{\infty} \phi(L') d \log_{10} L'
    \label{abmatchsubmm}
\end{equation}
to find $L(M,z = 0)$, which is then inserted into the framework above and propagated to high redshifts using an appropriate functional form, whose parameters are matched to the observations. 

The CO luminosity from galaxy surveys is usually measured in units of K km/s pc$^2$. This quantity is related to  the observed flux density of CO, and its linewidth, by the relation \cite{solomon2006}:
\begin{equation}
 L_{\rm CO} = 3.3 \times 10^{13} S \Delta v (1 + z)^{-3} \nu_{\rm obs}^{-2} D_L^2 \ \rm{K \  km/s \  pc}^2
\end{equation} 
with $D_L$ being the luminosity distance to the source in Gpc, $\nu_{\rm obs}$ the observed frequency in GHz, $S$ the flux density  in Jy, and $\Delta v$ the velocity width  in km/s. For the case of CO, a double power law form relating $L_{\rm CO}$ to halo mass is found to be a good fit to the data at low and high redshifts:
\begin{equation}
    L_{\rm CO} (M, z) = 2N(z) M [(M/M_1(z))^{-b(z)} + (M/M_1(z))^{y(z)}]^{-1}
\end{equation}
with the best fitting parameters and their uncertainties given in Table \ref{table:constraints}. This form is identical to the corresponding form of the abundance matched stellar mass to halo mass relation, found in data-driven approaches \cite{behroozi2010, behroozi2019, moster2010} and summarized in Table \ref{table:constraints}. 

For the case of [CII], a power law with an exponential cutoff matches the low-redshift data well, and the evolution to high redshifts is assumed to follow that of the star formation rate (since [CII] is found to be highly correlated with the star-formation rate, e.g., \cite{knudsen2016}) which is fitted utilizing a data driven procedure in \cite{behroozi2013, behroozi2019}. The relation is thus expressed as
\begin{equation}
 L_{\rm CII}(M,z) = \displaystyle{\left(\frac{M}{M_1}\right)}^{\beta} \exp(-N_1/M) \displaystyle{\left(\frac{(1+z)^{2.7}}{1 + [(1+z)/2.9)]^{5.6}} \right)^{\alpha}}   
\end{equation}
with the best fitting parameters  summarized in Table \ref{table:constraints}.
For [OIII], a fit to available observations of individual galaxies at high redshifts \cite{harikane2020} leads to a parametrization of the luminosity directly in terms of the star formation rate (Table \ref{table:constraints}).

 The advantages of using the above data-driven relations are manifold. A special feature of baryonic line emission power spectra (which can be seen from expressions like \eq{onehalo} and \eq{twohalo}) is that they
depend on the underlying cosmology as well as on the astrophysics of the  systems, and hence can offer constraints on both these aspects. The astrophysics acts as an effective ‘systematic’ uncertainty when making cosmological predictions from such surveys. At the same time, the astrophysical parameters themselves contain valuable information about the role of gas in galaxy formation and evolution. Using the latest available data to constrain the parameters of this framework therefore allows us to  precisely separate both these aspects. Furthermore, the models' computational simplicity allows us to include the effects of several additional parameters, easily vary the physics and cosmology to explore different scenarios, and to impose the most realistic priors (by definition, based on all the data available today) on the astrophysics conveniently within a Fisher matrix analysis. This has advantages for both cosmological and theoretical physics avenues, as it easily accounts for extensions beyond $\Lambda$CDM and is thus uniquely suited to extract cosmological constraints from baryonic data  and make predictions for future surveys \cite{hparaa2019}. We describe how this is done in the forthcoming sections.

\section{Constraining the baryons}
\label{s:constraints}

Using the framework described in the previous sections, one can now constrain the free parameters of the baryonic models by using the available astrophysical data. This is done by from the following datasets: (i) the galaxy and Damped Lyman Alpha system observations for \HI\ \cite{hptrcar2015}, fitted together using a Markov Chain Monte Carlo framework \cite{hparaa2017}, (ii) the galaxy surveys at low redshifts (\cite{keres2003} and \cite{hemmati2017}) coupled to high-redshift intensity mapping constraints (\cite{pullen2018} and \cite{keating2016})  for CO and [CII], and (iii) observations of [OIII] at high redshifts ($z \sim 6-9$) from individual galaxies \cite{harikane2020}, (iv) stellar mass to halo mass \cite{behroozi2019} and black hole mass to bulge mass observations \cite{ferrarese2002, kormendy2013, wyithe2002}.
This leads to the constraints on the parameters as given by Table \ref{table:constraints}.

\begin{longtable}{l|l|l}
\caption{Data-driven approaches for constructing the baryon - dark matter halo mass relations over a range of redshifts.  The columns list the
 technique, parameter(s) constrained, the mean
 redshift/redshift range, and the reference in the literature for each.} 
\\
\hline
&&\\
\tablehead
{\hline} Relation   & Redshifts & Reference \\
&&\\
\hline\hline
 & & \\
  {\bf \large HI Mass - halo mass}          &  & \\
  {\underline{Fitting function:}} &  &  \\
  & &  \\        
      $M_{\rm HI} (M) = \alpha f_{\rm H,c} M \left(M/10^{11} h^{-1} M_{\odot}\right)^{\beta} \exp\left[-\left(v_{c,0}/v_c(M)\right)^3\right]$  & $z \sim 0-5$ & Ref. \cite{hparaa2017}  \\ 
 $\rho_{\rm HI} (r) =\rho_0 \exp(-r/r_s)$; & & https://github.com/ \\ 
 & & JurekBauer/axion21cmIM\\
 $c_{\rm HI}(M,z) \equiv \displaystyle{\frac{R_v}{r_s}} = c_{\rm HI, 0} \displaystyle{\left(\frac{M}{10^{11} M_{\odot}}\right)^{-0.109}} 4/(1+z)^{\gamma}$ &  &  \\ 
  $v_c(M) = \displaystyle{\left(\frac{G M}{R_v(M)}\right)^{1/2}}$; $R_v(M)$ follows \eq{virialradius}. & & \\
  & & \\
  $f_{\rm H,c}$ follows \eq{fhc} & & \\
&&\\
{\underline{Parameter values:}} & & \\
&&\\
$c_{\rm HI,0} = 28.65 \pm 1.76$   &  &                                                           
\\
$\alpha = 0.09 \pm 0.01$  &  &  
\\
log$_{10} (v_{c,0}/\rm km \ s^{-1}) = 1.56 \pm 0.04$      &  &                                                           
\\
$\beta = -0.58 \pm 0.06$       &  &                                                                
\\
$\gamma = 1.45 \pm  0.04$      &  &  
\\
 \hline
 &&\\
     {\large {\bf CO Luminosity - Halo mass}}  & &    \\
      {\underline{Fitting function:}} &  & \\
     $L_{\rm CO} (M, z) = 2N(z) M [(M/M_1(z))^{-b(z)} + (M/M_1(z))^{y(z)}]^{-1}$ &  & \\
      & $z \sim 0-5$ & Ref. \cite{hpco} \\ 
       & & https://github.com/\\
      & & georgestein/limlam\_mocker  \\
     $\log M_1(z) = \log M_{10} + M_{11}z/(z + 1)$&& \\
$N(z) = N_{10} + N_{11}z/(z + 1)$&& \\
$b(z) = b_{10} + b_{11}z/(z + 1)$&& \\
$y(z) = y_{10} +  y_{11}z/(z + 1)$&& \\
& & \\
  {\underline{Parameter values:}} &  & \\
  &&\\
$M_{10}  = (4.17 \pm 2.03) \times 10^{12} M_{\odot}; \ M_{11} = (-1.17 \pm 0.85)$ &&\\
 $N_{10} = 0.0033 \pm 0.016 \ {\rm K \ km/s  \ pc}^2 M_{\odot}^{-1}; \ N_{11} = (0.04 \pm 0.03)$; && \\
 $b_{10} = 0.95 \pm 0.46, \  b_{11} = 0.48 \pm 0.35$; &&\\
 $y_{10} = 0.66 \pm 0.32, \  y_{11} = -0.33 \pm 0.24$ & & \\
 \hline
 & & \\
     {\large {\bf [CII] Luminosity - Halo mass} } &   &  \\
      {\underline{Fitting function:}} &  & \\
     && \\
     $L_{\rm CII}(M,z) = \displaystyle{\left(\frac{M}{M_1}\right)}^{\beta} \exp(-N_1/M) \displaystyle{\left(\frac{(1+z)^{2.7}}{1 + [(1+z)/2.9)]^{5.6}} \right)^{\alpha}} $ & & \\ 
      & $z \sim 0-7$ &  Ref. \cite{hpco} \\ 
  {\underline{Parameter values:}} &  & \\
$M_1 = (2.39 \pm 1.86) \times 10^{-5}$ &&\\
$N_1 = (4.19 \pm 3.27) \times 10^{11}$; && \\
 $\beta = 0.49 \pm 0.38$ &&\\
 &&\\
 \hline
 &&\\
    {\large {\bf [OIII] Luminosity - Halo mass}}  &   &  \\
    & & \\
       {\underline{Fitting function:}} &  & \\
      &&\\
     $\log\displaystyle{\left(\frac{L_{\rm OIII}}{L_{\odot}}\right)} = 0.97 
\times \log \displaystyle{\frac{\rm SFR}{[M_{\odot} \text{yr}^{-1}]}} + 7.4 $ \  & & \\ 
&&\\
${\rm SFR}(M)$ follows Ref. \cite{behroozi2019}   & $z \sim 6-9$ &  Refs. \cite{behroozi2019, harikane2020} \\
&&\\
      \hline
      &&\\
 {\large {\bf  Stellar Mass - Halo mass}}  &     \\
      & & \\
       {\underline{Fitting function:}} &  & \\
     &&\\
    $\log_{10}\displaystyle{\left(\frac{M_\ast}{M_1}\right)} =  \, \epsilon - \log_{10}\left(10^{-\alpha x} + 10^{-\beta x}\right) + \gamma\exp\left[-0.5\left(\frac{x}{\delta}\right)^2\right]$ & & \\
$x \equiv  \log_{10} \displaystyle{\left(\frac{M_\mathrm{peak}}{M_1}\right)}$ \ & $z \sim 0-10$ &  Ref. \cite{behroozi2019} \\ 
&&\\
      \hline
     &&\\
   {\large {\bf   Black hole Mass - Halo mass}}  &   &  \\
       & & \\
       {\underline{Fitting function:}} &  & \\
      &&\\
     $M_{\rm BH} (M) = M \epsilon_0 \displaystyle{\left(\frac{M}{10^{12} M_{\odot}}\right)^{\gamma/3 - 
1}} \left(\frac{\Delta_v \Omega_m h^2}{18 \pi^2}\right)^{\gamma/6} 
(1+z)^{\gamma/2} $ & & \\
& & \\
       {\underline{Parameter values:}} &  & \\
       $\epsilon_0 \sim 10^{-5}$ & $z \sim 0-8$ & Ref. \cite{wyithe2002}  \\
 $\gamma \sim 4.5$ &  &  
\label{table:constraints}
\\
\hline
\end{longtable}

\section{Learning cosmology from the baryons}\label{s:forecasts}
Over the past decade, the standard model of cosmology (i.e. $\Lambda$CDM, described in Sec. \ref{sec:astrocosmo}), has become fairly well established, with the latest constraints approaching sub-percent levels of accuracy in the measurement of cosmological parameters. However, from a theoretical point of view, several outstanding questions remain to be answered in the context of this model, of which the chief ones are: i) the nature of the late-time accelerated expansion of the Universe,  ii) the mechanism responsible for generating the primordial perturbations that led to structure formation, and (iii) the nature of dark matter. The first phenomenon is often attributed to dark energy, which behaves like a fluid component having negative pressure. Most observations are consistent with dark energy being a cosmological constant as defined in Sec. \ref{sec:astrocosmo}, however, the breakdown of the general theory of relativity on cosmological scales (also known as modifications of gravity) may also explain the observed cosmic acceleration. Modified gravity theories thus have a strong significance in understanding the nature of dark energy.  

The generation of primordial perturbations is believed to be related to an early inflationary epoch of the Universe. The detection of primordial non-Gaussianity, which is characterized by a nonzero value of the parameter $f_{\rm NL}$, places stringent constraints on theories of inflation, since standard single-field slow-roll inflationary models predict negligible non-Gaussianity (e.g., \cite{Maldacena:2002vr}), while non-standard scenarios allow for larger amounts. So far, searches for primordial non-Gaussianity have been through anisotropies in the Cosmic Microwave Background (CMB), or from the clustering of galaxies (e.g. \cite{Giannantonio:2013uqa, castorina2019}).

We are well-placed to study the impact of astrophysics on cosmological forecasts using the halo model treatments developed in the preceding sections. To do so, it is most convenient to \cite{hparaa2019} use a Fisher matrix technique, which enables realistic priors on the astrophysics imposed from the halo model framework. For large area sky surveys (such as those with \HI), the power spectra defined in Sec. \ref{s:HIhalomodel} are first converted into their projected \textit{angular} forms, denoted by $C_{\ell}(z)$ which represent
the observable, on-sky quantities in cosmological mapping surveys. The angular power spectrum can be used to perform a
tomographic analysis of clustering in multiple redshift bins without the 
assumption of an underlying cosmological model, e.g., \cite{seehars2016}. The expression for $C_{\ell}(z)$  can be related to the power spectra defined in Sec. \ref{s:HIhalomodel} and Sec. \ref{s:submmhalomodel} by using the  Limber approximation (accurate to within 1\% for scales above $\ell \sim 10$; e.g. Ref.\cite{limber1953}) that makes use of the angular window function, $W_{\rm HI}(z)$ of the survey:
\begin{equation}
C_{\ell} (z) = \frac{1}{c} \int dz  \frac{{W_{\rm HI}}(z)^2 
H(z)}{R(z)^2} 
P_{\rm HI} [\ell/R(z), z]
\label{cllimber}
\end{equation}
where $H(z)$ is the Hubble parameter at redshift $z$, and $R(z)$ is the comoving distance to redshift $z$. The angular window function is usually assumed to have a top-hat form in redshift space, though other choices are possible.
From the above angular power spectrum, and given an experimental configuration, 
a Fisher forecasting formalism can be used to place constraints on the 
 cosmological [e.g., $\{h, \Omega_m, n_s, \Omega_b, \sigma_8\}$]  and 
astrophysical [e.g., $\{c_{\rm HI}, \alpha, \beta, \gamma, v_{\rm c,0}\}$] 
parameters, which are generically denoted by $A$. The Fisher matrix element 
corresponding to the parameters $\{A,B\}$ and at a redshift bin centred at $z_i$ is then defined as:
\begin{equation}
F_{AB} (z_i) = \sum_{\ell < \ell_{\rm max}} \frac{\partial_A C_\ell (z_i)
\partial_B C_\ell(z_i)}{\left[\Delta C_\ell(z_i)\right]^2},
\end{equation}
where $\partial_A$ is the partial derivative of $C_{\ell}$ with respect to 
$A$. In the above expression, the standard deviation, $\Delta C_{\ell} (z_i)$ is defined 
in terms of the noise of the experiment, $N_{\ell}$ and the sky 
coverage of the survey, $f_{\rm sky}$:
\begin{equation}
\Delta C_\ell = \sqrt{\frac{2}{(2 \ell + 1)f_{\rm sky}}} 
\left(C_\ell + N_\ell\right),\label{eq:Delta_Cl}
\end{equation}
 The full Fisher matrix for an experiment is constructed by summing the individual Fisher matrices in each of the $z$-bins
in the redshift range covered by the survey: 
\begin{equation}
 \mathbf{F}_{AB} = \sum_{z_i} F_{AB}(z_i)
 \label{fisher}
\end{equation}

The above treatment for cosmological forecasting  exactly parallels the corresponding one for the CMB, e.g., \cite{bond1997}, with the additional appearance of the baryonic gas parameters. Note that the Fisher matrix approach assumes that the individual parameter likelihoods are approximately Gaussian, which may not always be the case if the parameters are strongly degenerate. In the present case, however, it is found \cite{hparaa2019} that a full Markov Chain Monte Carlo treatment leads to negligible differences from that obtained by the Fisher matrix technique.

For sub-millimetre surveys that typically cover only a few square degrees of the sky, the three-dimensional power spectrum $P_{\rm submm}(k)$ defined in \eq{submmpower} is directly used to compute the Fisher matrix, which is written as:
\begin{equation}
F_{AB} (z_i) = \sum_{k < k_{\rm max}} \frac{\partial_A P_{\rm submm}(k, z_i)
\partial_B P_{\rm submm}(k,z_i)}{\left[\sigma_P(k,z_i)\right]^2},
\end{equation}
The variance of the power spectrum is denoted by $\sigma_P^2(k, z_i)$ and is computed as:
\begin{equation}
    \sigma_P^2 = \frac{(P_{\rm submm}(k, z_i) + P_{\rm N})^2}{{N_{\rm modes}(k)}}
    \label{varianceauto}
\end{equation}
where $P_{\rm N}$ is the noise power spectrum, analogous to $N_{\ell}$ in \eq{eq:Delta_Cl} above, and $N_{\rm modes}$ is the number of Fourier modes probed by the survey, defined as:
\begin{equation}
    N_{\rm modes} = 2 \pi k^2 \Delta k \frac{V_{\rm surv}}{(2 \pi)^3} 
\end{equation}
in which $V_{\rm surv}$ is the volume of the survey and $\Delta k$ denotes the spacing of the $k$-bins.

To complete the treatment, we need to consider the experimental noise in various configurations which defines  $N_{\ell}$ and $P_{\rm N}$ above.  The way the noise is calculated differs according to the configuration in which the facility is constructed and used. For HI, the experimental configurations can be roughly divided into three categories: (i) single dish telescopes, such as the Five Hundred Metre Aperture Spherical Telescope (FAST, \cite{smoot2017}), BAO in Neutral Gas Observations (BINGO, \cite{battye2012}), and the Green Bank Telescope (GBT) and  (ii) dish interferometers, such as the TianLai \cite{chen2012}, the Square Kilometre Array (SKA) and its pathfinders, like the Meer-Karoo Radio Telescope (MeerKAT, \cite{santos2017}) and the Australian SKA Pathfinder (ASKAP, \cite{johnston2008}), the Hydrogen Intensity and Real-time Analysis eXperiment (HIRAX, \cite{newburgh2016}) and cylindrical configurations, like the Canadian Hydrogen Intensity Mapping Experiment (CHIME, \cite{bandura2014}) and the planned CHORD (Canadian Hydrogen Observatory and Radio transient Detector) \cite{liu2019} facilities. For submillimetre lines, planned or already-fielded instruments include interferometer arrays, such as the Sunyaev-Zeldovich Array (SZA) \cite{keating2016}, as well as single dish facilities such as COMAP\footnote{http://comap.caltech.edu} and CCAT-p \cite{terry2019, parshley2018}, in addition to balloon-based experiments such as EXCLAIM \cite{exclaimpaper2020}. A summary of the noise expressions for the various configurations relevant to these experiments is provided in Table \ref{table:noise}.

\begin{longtable}{l|l|l}
\caption{Expressions for the noise terms $N_{\ell}$ and $P_{\rm N}$ used in the Fisher information matrix to constrain cosmological and astrophysical parameters with the intensity mapping technique at various redshifts and using various tracers. Key to symbols: $\Delta \nu$ : observed frequency interval;  $T_{\rm sys}$: system temperature, $f_{\rm sky}$ : sky fraction covered, $\lambda_{\rm obs}$ : observed wavelength, $\theta_{\rm beam}$ : telescope beam size, $\bar{T}$ : mean temperature of hydrogen, $D_{\rm dish}$ : the dish size, $N_{\rm dish}$ : the number of dishes, $\nu$ : the observed frequency, $W_{\rm cyl}$ : width of the cylinder, the FoV : Field of View, $A_{\rm eff}$ : effective area,
$t_{\rm pix}$ : time observing per pixel, $n_{\rm base}$ : number of baselines, $n_{\rm pol}$ : number of polarization directions, $N_{\rm beam}$ : number of beams, $D_{\rm max/,min}$: maximum and minimum baselines of the interferometer configuration,
   $V_{\rm vox}$: volume of the `voxel' (volume pixel) in submillimetre surveys, $N_{\rm det}$: number of detectors, $\sigma_{\rm N}$: detector noise per voxel,  $t_{\rm vox}$: time spent observing the voxel, and $t_{\rm tot}$ : total observational time.}
\\
\hline
&&\\
\tablehead
{\hline} Experiment   & Noise expression & Reference \\
&&\\
\hline\hline
 & & \\
 {\large \bf  (i) HI single dish}         &   $N_\ell^{\rm HI,dish} = {W _{\rm beam}^2(\ell)}/{2N_{\rm dish}  t_{\rm pix} \Delta \nu} \left({T_{\rm sys}}/{\bar{T}}\right)^2\left({\lambda_{\rm obs}}/{D_{\rm dish}}\right)^2$ & \\
&  & Ref. \cite{camera2020} \\
& $W _{\rm beam}^2(\ell)=\exp\left[{\ell(\ell+1)\theta_{\rm beam}^2}/{8\ln2}\right]$  & \\
   {\large \bf (ii) HI interferometer} & & \\
  {\large \bf in single dish mode}  & $N_\ell^{\rm HI,intSD} = 4 \pi f_{\rm sky} T_{\rm sys}^2/(2 N_{\rm dish} t_{\rm tot} \Delta \nu)$ & \\
  & & \\
 &   $T_{\rm sys} = 25 + 60 \left({300 \ {\rm MHz}}/{\nu}\right)^{2.5}$ & Refs. \cite{knox1995,bull2014,ballardini2019,bauer2021}  \\
{\large \bf (iii) HI cylindrical} & $N_\ell^{\rm HI,cyl} = \displaystyle{\frac{4 \pi f _{\rm sky}}{{\rm FoV} n_{\rm base}(u) n_{\rm pol} N_{\rm beam}  t_{\rm tot} \Delta \nu}} \displaystyle{\left(\frac{\lambda_{\rm obs}^2}{A_{\rm eff}}\right)^2}\displaystyle{\left(\frac{T_{\rm sys}}{\bar{T}}\right)^2};$ &   \\
   &  &  \\
   & & \\
 & ${\rm FoV} \approx \pi/2 \lambda_{\rm obs}/W_{\rm cyl}$; & \\
 & & \\
 & $\Delta \nu = \nu_{\rm HI} \Delta z/(1+z)^2$ & Refs. \cite{camera2020, newburgh2014, jalilvand2019, obuljen2018}  \\
 & & \\
 {\large \bf (iv)  HI interferometer} & $N_{\ell}^{\rm HI,  int} = T_{\rm sys}^2 {\rm FoV}^2/(T_b^2 n_{\rm pol} n(u = \ell/2\pi)  t_{\rm tot} \Delta \nu)$ & \\
 & & \\
 &  $n(u) = N_{\rm dish} (N_{\rm dish} - 1)/(2 \pi (u_{\rm max}^2 - u_{\rm min}^2))$ & \\
 & & \\
 & FoV $= \lambda^2/D_{\rm dish}^2$; \ $u_{\rm max/min} = D_{\rm max,min}/\lambda$ & Refs. \cite{bauer2021, pourtsidou2016, bull2014}  \\
 & & \\
{\large \bf (v) CO} & $P_{\rm N} = V_{\rm vox} \sigma^2_{\rm vox}$ &  \\
& & \\
 & $\sigma_{\rm vox} = T_{\rm sys}/(N_{\rm det} \Delta \nu t_{\rm vox})^{1/2}$ & Ref. \cite{liu2021} \\
 & & \\
{\large \bf   (vi) CII/OIII IM} &  $P_{\rm N} = V_{\rm vox} \sigma_{\rm N}^2/t_{\rm vox}$ & Ref. \cite{hpcii2019} \\
& &
 \label{table:noise}
 \\
\hline
\end{longtable}
Given the Fisher matrix in \eq{fisher}, we can then compute the standard deviation in the measurement of $A$ for the two cases of fixing and marginalizing over the remaining parameters, as:
\begin{equation}
   \sigma^2_{A, \rm fixed} = (\mathbf{F}_{AA})^{-1}; \ \sigma^2_{A, \rm marg} = (\mathbf{F}^{-1})_{AA};
\end{equation}
The Fisher matrix treatment thus captures the effects of `astrophysical degradation' in the precision of cosmological parameter forecasts, and their evolution with redshifts for different experiment combinations. It was found that, in the case of \HI, the astrophysical degradation was largely mitigated by our prior information coming from the present knowledge of the astrophysics \cite{hparaa2019}. It also revealed an important robustness feature of the halo model framework in Sec. \ref{s:HIhalomodel}: that physically motivated extensions to the halo model did not cause significant changes in the forecasts, which was important to consolidate the utility of the halo model framework for constraining theories of fundamental physics.

It is also important to assess the influence of astrophysical uncertainties on the \textit{accuracy} of cosmological parameter forecasts \cite{camera2020}. This can be addressed  by  employing the nested likelihoods framework \cite{Heavens:2007ka}, which is a measurement of how much the uncertainty in our astrophysical knowledge causes a bias in the forecasted cosmological parameters. Specifically, given the Fisher matrix in \eq{fisher}, we
split the space of parameters into two 
subsets: one containing the parameters of interest and the other containing the parameters
deemed `nuisance' or systematic for the analysis under consideration. When constraining cosmology with baryons,
these two sets could represent `cosmological' and `astrophysical' 
parameters respectively. The bias on a given cosmological parameter $A$, 
denoted by $b_{A}$, is then computed as:
\begin{equation}
b_{A} = \delta \mu F_{B\mu}\left(\mathbf 
F^{-1}\right)_{AB}.\label{eq:bias}
\end{equation}
Here, $\mathbf F^{-1}$ is identical to \eq{fisher} and represents the full Fisher matrix of astrophysical and 
cosmological parameters, and $F_{B\mu}$ stands for the particular sub-matrix mixing 
cosmological and astrophysical parameters. The term $\delta \mu$ denotes 
the vector of the shifts between the fiducial and true values of the astrophysical parameters $\mu$:
\begin{equation}
\delta \mu=\mu^{\rm fid}-\mu^{\rm true}.\label{eq:shift}
\end{equation}

\begin{figure}
\centering
\includegraphics[width = 0.9\textwidth]{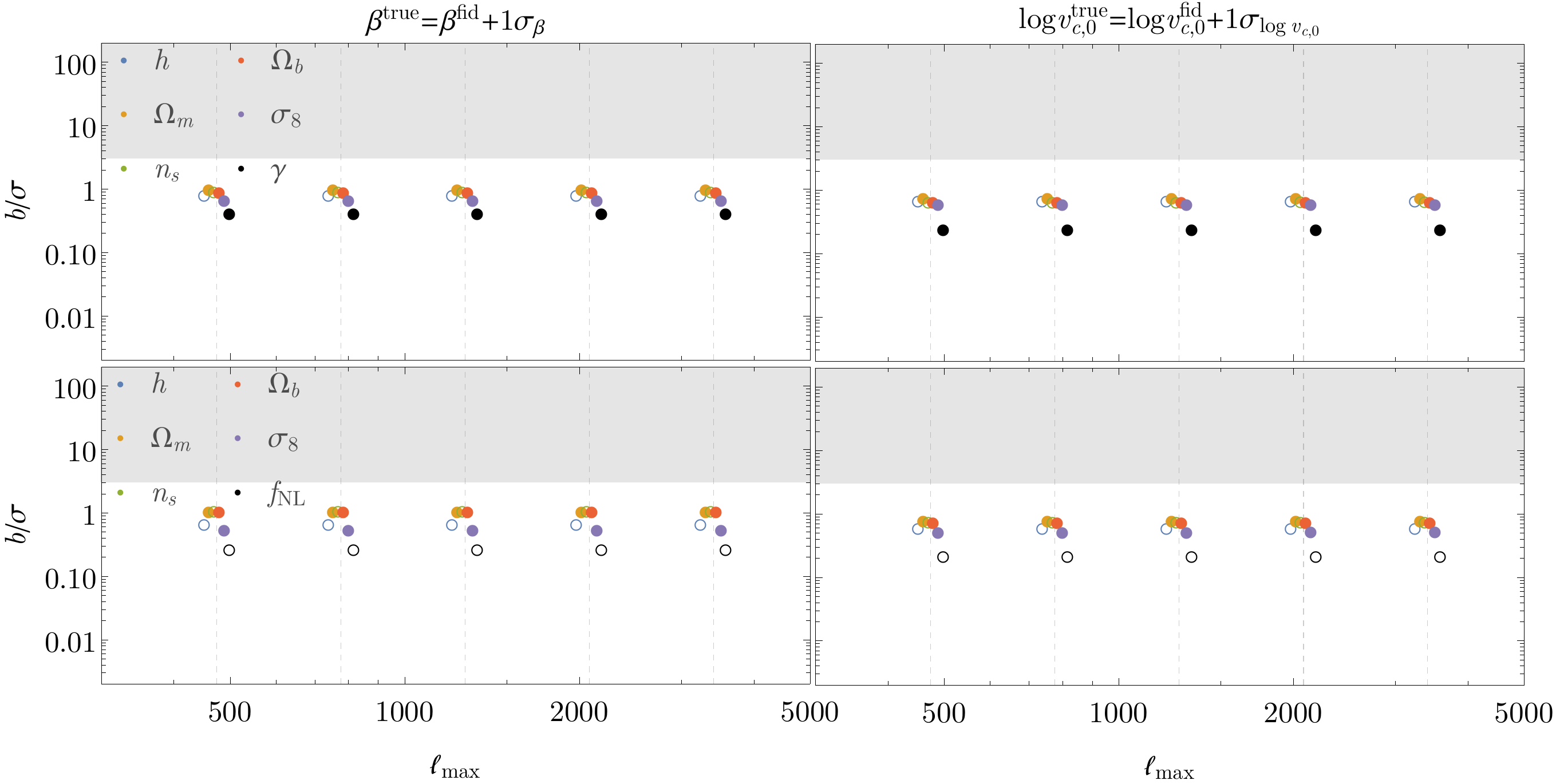}
\caption{Relative bias $b/\sigma$ on cosmological parameters, including those beyond the standard $\Lambda$CDM 
framework,  with a SKA I MID-like experimental configuration, obtained on shifting either 
astrophysical parameter, $\beta$ (left panels) or $\log v_{\rm c,0}$ (right panels), 
by $1 \sigma$ from its mean value given in Table \ref{table:constraints}. Top panels show the biases in
$\Lambda$CDM+$\gamma$ where $\gamma$ is a measure of modified gravity \cite{camera2020}, and lower panels show biases for those in $\Lambda$CDM+$f_{\rm NL}$. The empty 
(filled) circles indicate negative (positive) values of  biases. It can be seen that the bias values are well within 1$\sigma$ for all the parameters considered, including the $f_{\rm NL}$. Shaded areas (which are not reached by the results) indicate the range $b/\sigma > 3$, for which it is possible for the Gaussian approximation to the likelihood to become inaccurate. Figure taken from \cite{camera2020}. }
\label{fig:relbias1}
\end{figure}

This approach thus allows us to quantify the accuracy of cosmological forecasts obtainable from baryonic surveys. The relative bias on various cosmological parameters from a SKAI-MID like survey, obtained by shifting the parameters $v_{c,0}$ and $\beta$ from their fiducial values in Table \ref{table:constraints}, is plotted in Fig. \ref{fig:relbias1}.
As a bonus, it was found that \cite{camera2020} this technique leads to a powerful way of incorporating effects beyond the standard $\Lambda$CDM framework into the halo model. Specifically, we can use this method to investigate a non-zero primordial non-Gaussianity effect imprinted on the power spectrum, and also incorporate modified gravity scenarios which are an important test of Einstein’s general relativity. Encouragingly, it was found that the primordial non-Gaussianity is negligibly affected by astrophysical uncertainties as can be seen from Fig. \ref{fig:relbias1}, which promises an optimistic outlook for one of the strongest science cases for future intensity mapping experiments.

Dark matter (DM) is also a component of the $\Lambda$CDM cosmological model, however, its nature continues to be a mystery. The only dark matter candidate in the standard model of particle physics, the neutrino, is known to make up less than 1\% of the total DM abundance because its relativistic velocity makes it too “hot” to account for the observed structure formation (e.g., \cite{Alam:2016hwk}). Observations  favour the majority of the remaining DM being composed of a single species of cold, collisionless DM (CDM). One candidate for a significant fraction of DM are axions, which occur in  many extensions of the standard model \cite{PecceiQuinn1977, Weinberg1978}. Axions with masses $m_a \sim 10^{-22}$ eV, known as ``fuzzy DM" \cite{Hu:2000ax}, can make up a significant fraction of the DM, and furthermore have a host of interesting phenomenological consequences on galaxy formation (e.g. \cite{Arvanitaki_2010, Marsh_review2016, Niemeyer:2019aqm}). 

Ref. \cite{bauer2021} used the halo model for HI described in Sec. \ref{sec:analytical} to explore the effects of an axion subspecies of DM in the mass range $10^{-32}$ eV $\leq m_a \leq$ $10^{-22}$ eV on the HI power spectrum at $z \leq 6$. It was found that lighter axions introduce a scale-dependent feature even on linear scales due to the suppression of the matter power spectrum near the Jeans scale. For the first time, it was possible to forecast the bounds on the axion fraction of DM in the presence of astrophysical and model uncertainties, achievable with upcoming facilities such as the HIRAX and SKA. A compilation of the latest forecasted constraints on the above beyond-$\Lambda$ CDM parameters with future surveys mapping baryonic gas is provided in Table \ref{table:beyondlcdm}.

\begin{landscape}
\begin{longtable}{l|l|l}
\caption[]{Latest forecasted constraints on non-standard dark matter (e.g., \cite{khlopov2013}), tests of inflation and modified gravity with the future experiments tracing baryonic gas, primarily in the radio and sub-millimetre regimes.} 
\tablehead
{\hline Type of constraint  & Limit & Experiment/Reference \\ \hline\hline} 
\tabletail
{\hline \multicolumn{4}{r}{\textit{Continued on next page}}\\}
\tablelasttail{\hline} \\
\hline
{\bf Constraints on parameters describing non-cold dark matter} & &  \\
\hline
Warm dark matter particle mass & $m_{\rm WDM} = 4$ keV ruled out at $> 2 \sigma$ at & \\
&  $z \sim 5$ & SKAI-LOW \cite{carucci2015} \\
Effective parameter for & & \\
dark matter decays & $\Theta_{\chi} \approx 10^{-40}$ at $10^{-5}$ eV & HERA/HIRAX/CHIME \\
& & \cite{bernal2021} \\
Axion mass & $m_a \approx 10^{-22}$ eV, at 1\% & SKAI-MID + CMB \cite{bauer2021} \\
Particle to two-photons coupling of axion-like particles (ALPs) & $g_{a \gamma \gamma} \leq 10^{-11}$/GeV & SPHEREx/LSST \cite{shirasaki2021} \\
\hline
{\bf Primordial non-Gaussianity constraints}  & & \\
  \hline
Standard deviation of  & $\sigma(f_{\rm NL}) \sim 4.07$ & SKA \cite{gomes2020} \\
 non-Gaussianity parameter & $\sigma(f_{\rm NL}) < 1$ & PUMA \cite{karagiannis2020} \\
& $\sigma(f_{\rm NL}) \sim 10$ & COMAP \cite{liu2021} \\
& $\sigma(f_{\rm NL}) \sim 2-3$ & ngVLA  \\
& $\sigma(f_{\rm NL}) \sim 2-3$ & PIXIE/OST MRSS \\
& & \cite{dizgah2019}\\
& $\sigma(f_{\rm NL}) < 1$ & SKA1+Euclid-like+CMB Stage 4 \cite{ballardini2019a} \\
\hline
{\bf Constraints on modified gravity parameters}  & &  \\
 \hline
Effective gravitational strength; & &  \\
initial condition parameter of matter perturbations & $Y (G_{\rm eff}), \alpha < 1\%$ at $z \sim 6-11$ & SKA \cite{heneka2018} \\
Parameter modification of $f(R)$ gravity & $B_0 < 7 \times 10^{-5}$ & CMB + 21-cm \cite{hall2013} \\
Value of $f(R)$ field in background today & $|f_{R0}| < 9 \times 10^{-6}$ & 21 cm \cite{masui2010} \\
\hline
\label{table:beyondlcdm}
\end{longtable}
\end{landscape}

\begin{figure}
\centering
\includegraphics[width = 0.5\textwidth]{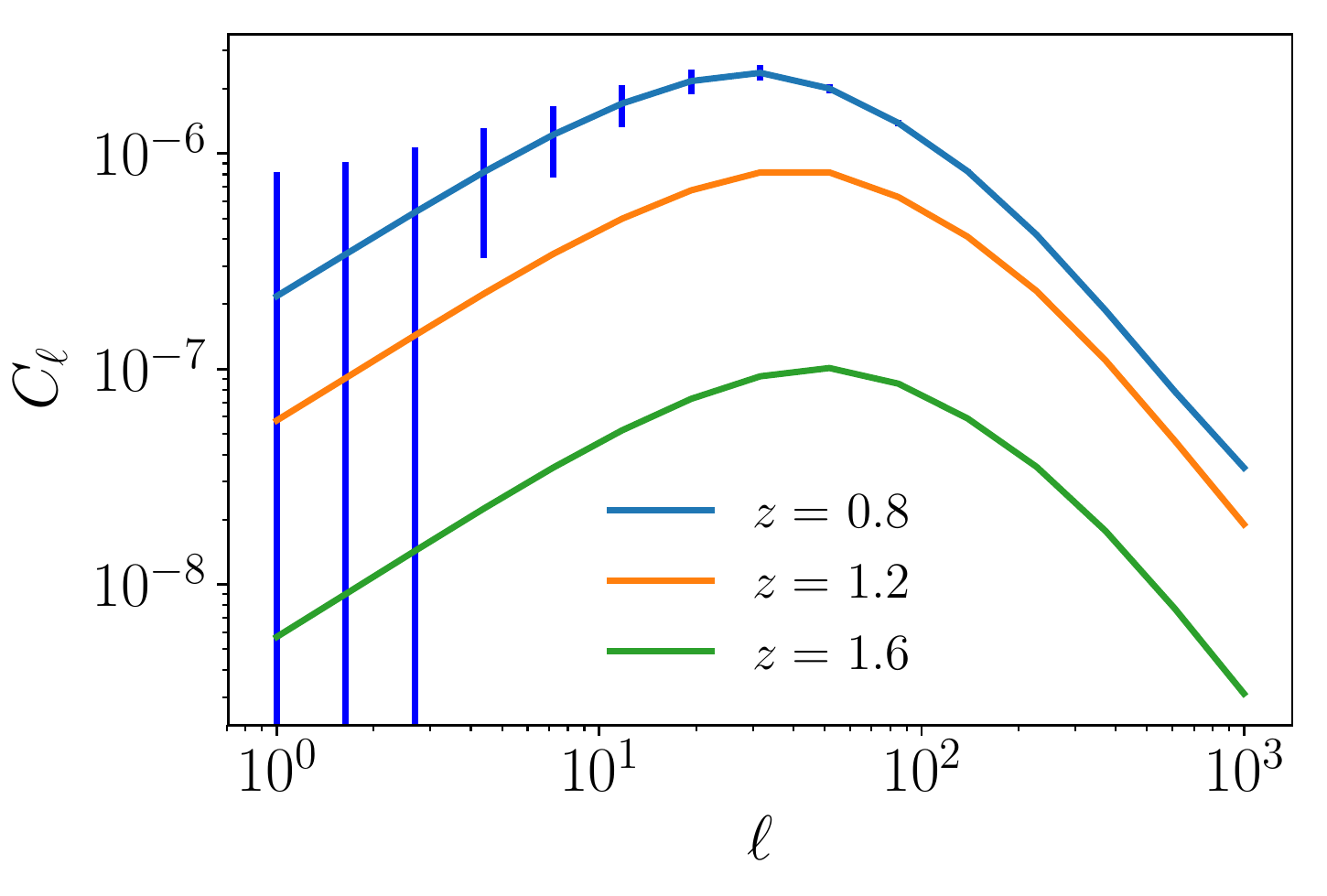}\includegraphics[width = 0.5\textwidth]{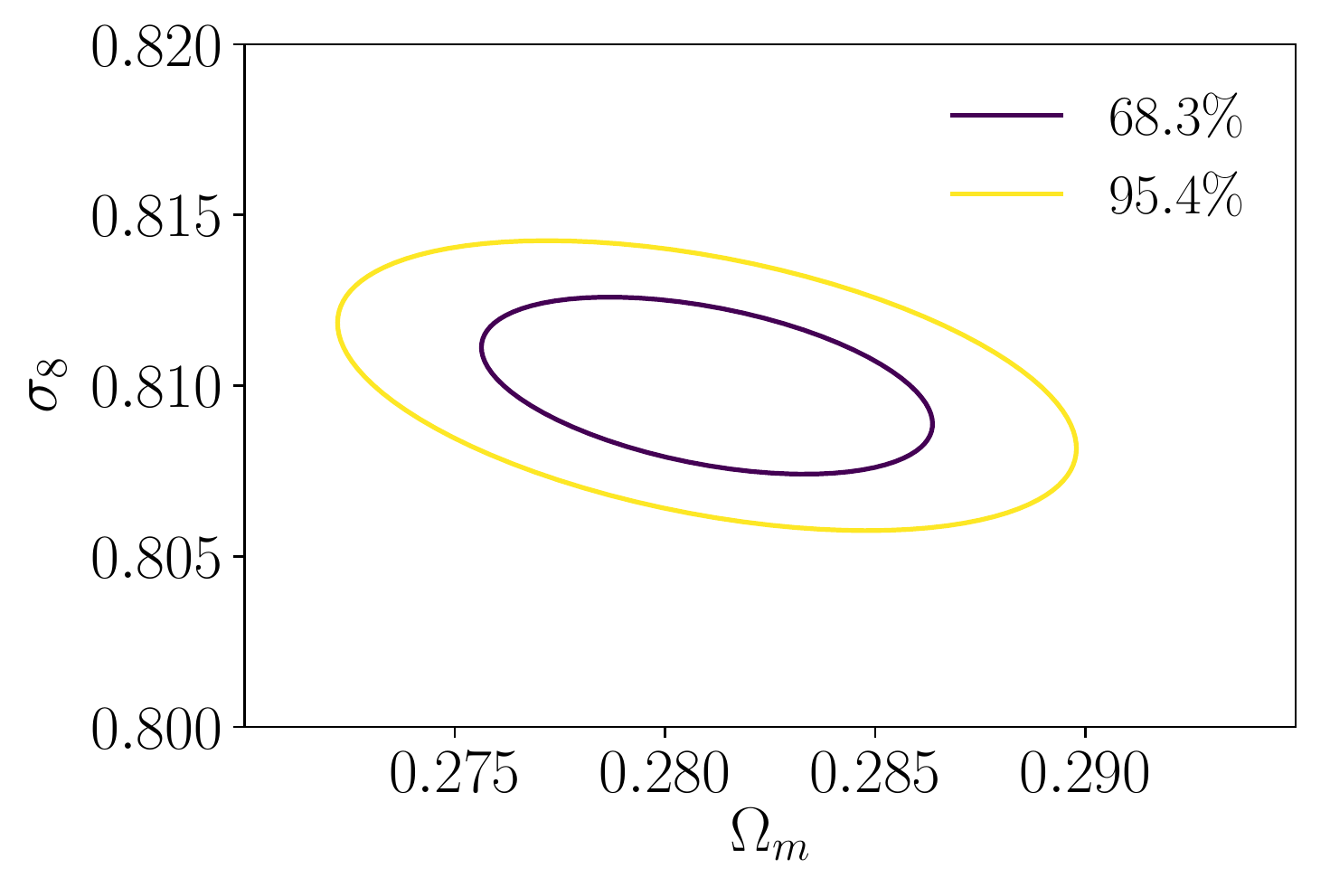}
\caption{\textit{Left panel:} Angular cross-correlation power spectrum computed from \eq{angcross} with a CHIME-DESI-like survey combination at redshifts 0.8, 1.2, and 1.6. The error bars denote the corresponding $\Delta C_{\ell}$ at the lowest redshift bin. Figure from \cite{hparaa2019}. \textit{Right panel:} Contours of cosmological parameters $\sigma_8$ and $\Omega_m$ with \textit{all} other parameters fixed in the cross-correlation.}
\label{fig:contour1}
\end{figure}

\section{The whole is greater than the sum of its parts}
\label{sec:crosscorrelations}

In this section, we describe how bringing together two (or more) different surveys leads to several advantages in the measurement of cosmological and and astrophysical parameters from baryonic tracers. Furthermore, it offers a direct route to extend the modelling frameworks towards the reionization regime, connecting up with future gravitational wave measurements to provide a holistic picture of cosmic dawn.   

\subsection{Gas-galaxy cross-correlations}

If we cross-correlate maps of gas emission (like HI) and galaxy surveys in the optical band, we can significantly improve our accuracy in the prediction of astrophysical parameters. This happens due to the foregrounds and systematics in the two surveys being mitigated in the cross-correlation between the two maps, thereby increasing the significance of the detection. The first detections of HI in intensity mapping took place using this approach \cite{Chang:2010jp, switzer13, masui13} with the HI observations from the Green Bank Telescope cross-correlated with galaxy data from the WiggleZ Dark Energy Survey probing $z \sim 1$, and more recently with galaxies from the eBOSS survey \cite{wolz2021}. Also, the cross-correlation of \HI\ intensity maps obtained from the Parkes telescope with 2dF galaxy Redshift Survey has recently been conducted at a lower redshift, $z \sim 0.08$ \cite{anderson2018}. 

\begin{figure}
\centering
\includegraphics[width = \textwidth]{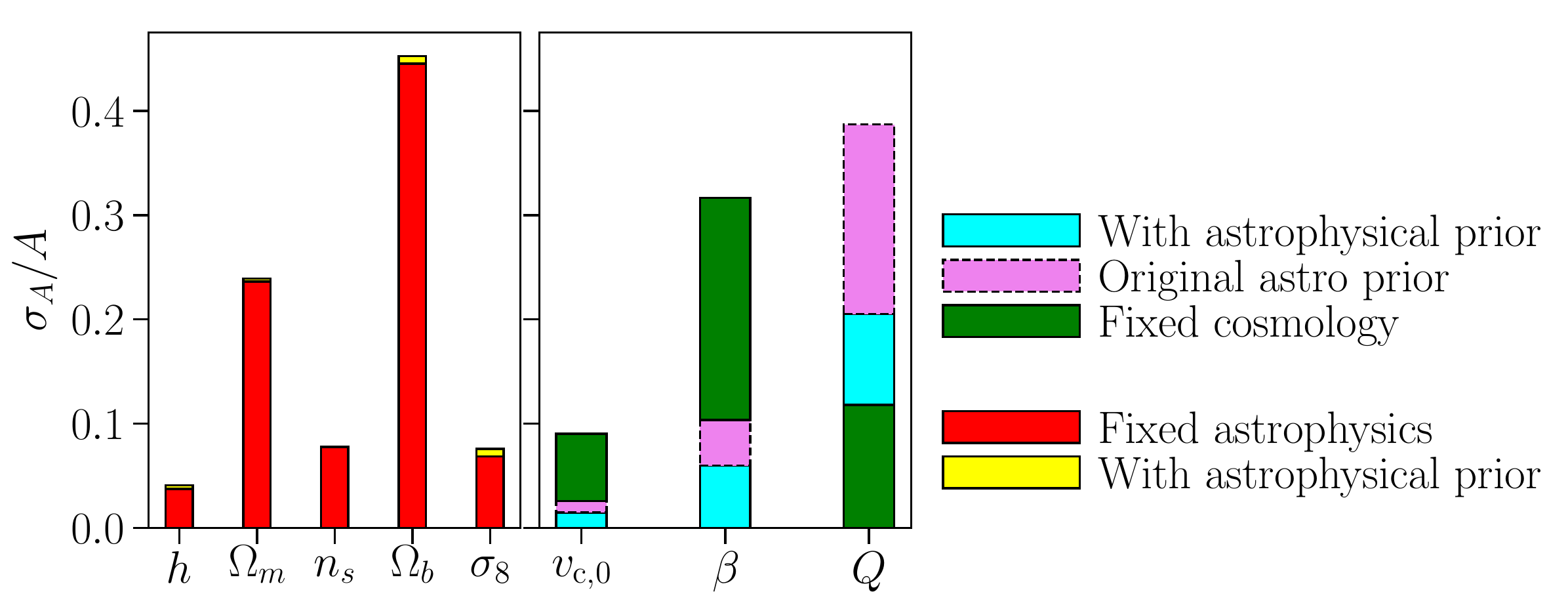}
\caption{Relative accuracy on forecasted cosmological and astrophysical parameters with a CHIME-DESI-like survey cross-correlation covering redshifts $0.8 < z < 1.6$. \textit{Left panel:} Constraints on the cosmological parameters $\{h, \Omega_m, n_s, \Omega_b, \sigma_8\}$ in a flat $\Lambda$CDM framework, for (i) fixed values of astrophysical parameters (red) and (ii) marginalizing over astrophysical parameters (yellow). \textit{Right panel:} Constraints on the HI-halo mass parameters ($v_{c,0}$ and $\beta$ from Table \ref{table:constraints}, and the $Q$ parameter which denotes the large scale galaxy bias \cite{cole2005}) for (i) fixed values of cosmological parameters (green), (ii) marginalizing over the cosmological parameters but adding an astrophysical prior (cyan). The extent of the astrophysical prior from current data is plotted as the violet band in each case.
 Figure from \cite{hparaa2019}.}
\label{fig:contour2}
\end{figure}

For a cross-correlation survey of \HI\ and galaxies, the observable angular power spectrum is modified from \eq{cllimber} to read (e.g., \cite{hpcrosscorr2020}):
\begin{equation}
C_{\ell, \times} = \frac{1}{c} \int dz \frac{{W_{\rm HI}(z) W_{\rm gal}(z)} 
H(z)}{R(z)^2} 
(P_{\rm HI} P_{\rm gal})^{1/2}
\label{angcross}
\end{equation}
which is illustrated in the left panel of Fig. \ref{fig:contour1} for a Canadian Hydrogen Intensity Mapping Experiment (CHIME)-like survey covering the redshift range 0.8-2.5, cross-correlated with a Dark Energy Spectroscopic Instrument (DESI, \cite{desi2016})-like Emission Line Galaxy (ELG) galaxy survey over the range $z \sim 0.6-1.8$. The angular power spectrum above contains both the \HI\ and galaxy linear power spectra, as well as the corresponding window functions  ($W_{\rm gal}$ can in general be different from $W_{\rm HI}$, and depends on the details of the selection function of the galaxy survey, e.g.,\cite{smail1995}). Such a cross-correlation promises much more stringent constraints (shown in the right panel of Fig. \ref{fig:contour1}) on the astrophysical and cosmological parameters than auto-correlation (correlating galaxies with galaxies), as already shown in several recent analyses,  e.g., cross-correlating a CHIME-like and DESI-like survey leads to an improvement by  factors of a few in the cosmological and astrophysical constraints when compared to those from the CHIME-like survey alone (Ref.\cite{hpcrosscorr2020}, shown in Fig.\ref{fig:contour2}). Notably, this improvement occurs in spite of the fact that the redshift coverage of the cross-correlation is only about half that of the CHIME-like autocorrelation survey, so this illustrates the extent to which adding the galaxy survey information helps to improve the cosmological constraints. 
 It was shown \cite{shi2020} using the halo model framework  described in Sec. \ref{sec:analytical} that the broadband BAO feature measured from the angular cross-correlation power spectra between a DESI  Emission Line Galaxy (ELG) survey and the 21 cm intensity maps measured from the TianLai survey\footnote{http://tianlai.bao.ac.cn/}, can be used to forecast a constraint on the angular diameter distance with a precision of 2.7\% over $0.775 < z < 1.03$, which is complementary to the BAO cosmic distance measured by galaxy-galaxy auto-correlation.
 
In the sub-millimetre regime, the cross-correlation prospects for the COMAP (CO Mapping Array Project) survey and the photometric COSMOS and spectroscopic HETDEX Lyman-alpha fields \cite{chung2019} showed that about 0.3\% accuracy in redshifts with greater than 0.0001 sources per cubic Mpc, with spectroscopic redshift determination should enable a CO-galaxy cross spectrum detection significance at least twice that of the CO auto spectrum (even if the CO survey covers only a few square degrees). This illustrates that cross-correlations with galaxy surveys can make significant improvements to the goals of future intensity mapping experiments in the submillimetre regime.

\begin{figure}
\begin{center}
\includegraphics[width = 0.45\textwidth]{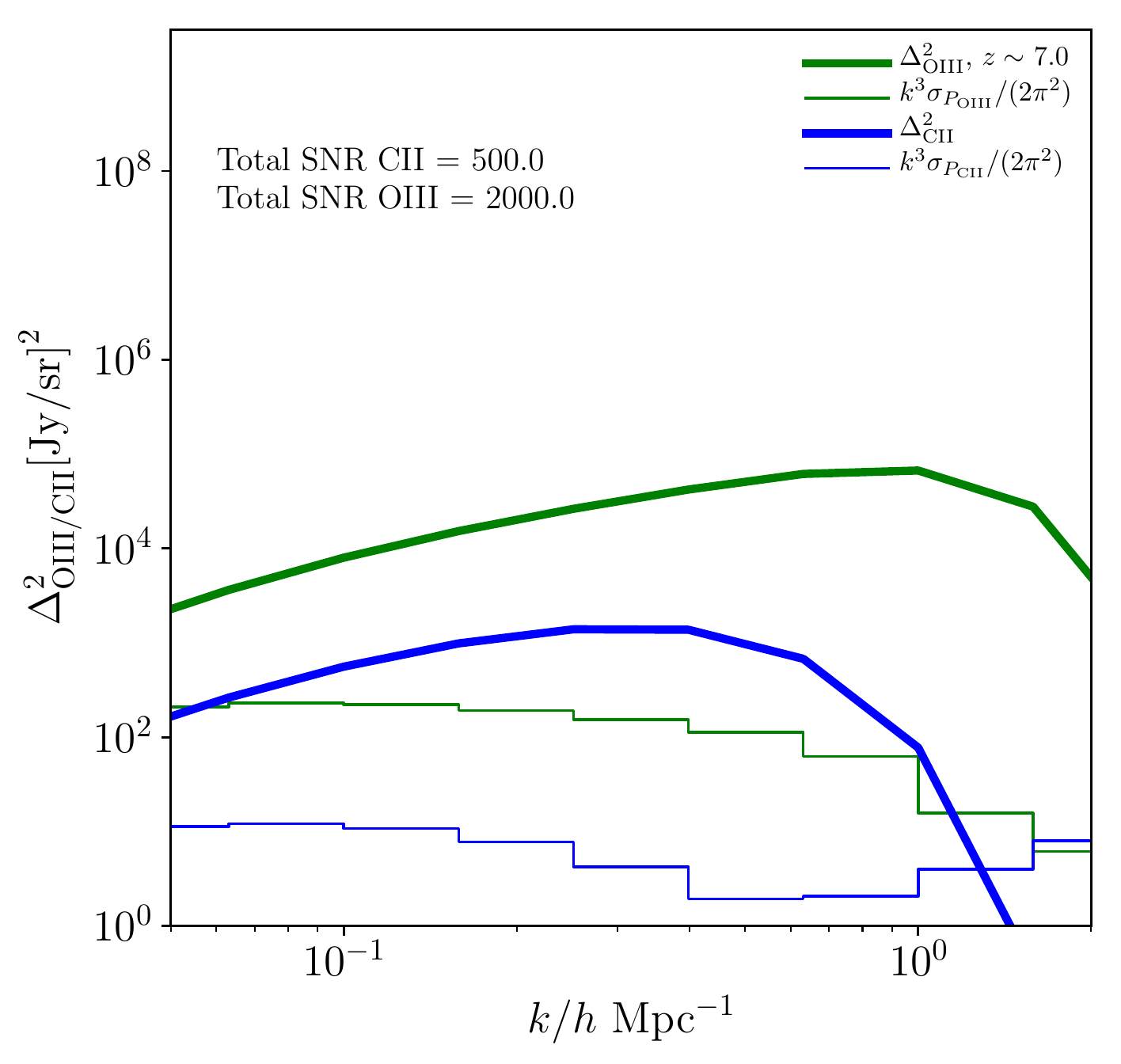} \includegraphics[width=0.45\textwidth]{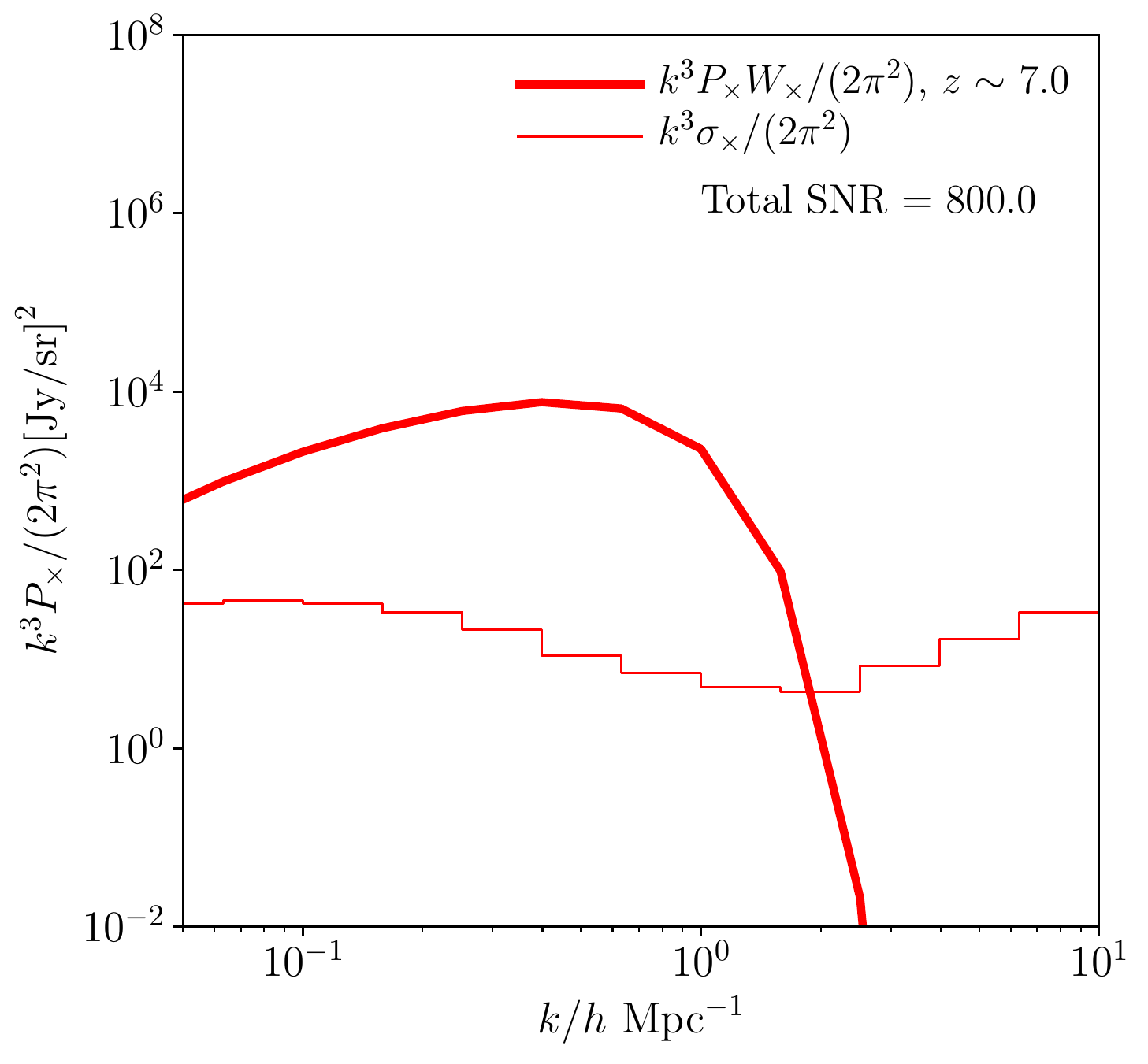}
\caption{\textit{Left panel}: Thick lines show the forecasts for the autocorrelation power spectra (computed using \eq{COpowspeclog} with a correction for the finite size of the telescope beam) of the [OIII]  88 $\mu$m transition (green) and the [CII] 158 $\mu$m transitions (blue) at $z \sim 7$. Overplotted in thin steps are the noise estimates (using \eq{varianceauto}) for a designed survey exploiting the potential of current architecture. \textit{Right panel:} Cross correlation of [OIII] 88 $\mu$m with [CII] 158 $\mu$m with the design configuration (red line), and the associated noise (thin steps). The total signal-to-noise is indicated in all cases. Figure adapted from \cite{hpoiii}.}
\label{fig:oiii}
\end{center}
\end{figure}

\subsection{Towards the reionization regime}
\label{s:estimator}

Thus far, we have considered the evolution of baryonic gas in the late-time universe, notably hydrogen and carbon inside galaxies and discussed the cosmological and physics constraints that can be extracted from their surveys. As we saw from Sec. \ref{sec:astrocosmo}, up to redshifts of about $z \leq 6$ or so, the neutral gas is primarily inside galaxies, with most of the intervening material being taken up by the ionized Lyman-$\alpha$ forest.

At higher redshifts ($z > 6-10$ or so), the situation is expected to be very different. The intergalactic hydrogen is predominantly neutral since the universe has not yet been reionized. Moreover, before virialized structures (stars, galaxies) form in large numbers, the baryons act as excellent tracers of the underlying dark matter.  The neutral hydrogen (the predominant baryon at high redshifts) is therefore a direct probe of the \textit{cosmic web}, and its mapping is thus expected to shed light on the process of structure formation itself.

The epoch of Cosmic Dawn, where the first stars and galaxies were born, signals the start of the second major phase transition of nearly all the normal matter in the universe, i.e., Cosmic Reionization which was introduced in Sec. \ref{sec:reion}. 
Reionization is characterized by the development of ionized regions (called ‘bubbles’) around the first luminous sources, and the end of reionization is marked by the complete overlap of these bubbles. The distribution of the bubble sizes leads to fluctuations in the neutral hydrogen density field, and thus in the signal observed with future 21 cm experiments. Hence, mapping the distribution of the ionized bubbles is crucial for modelling the observable signal. The bubble size distribution has been investigated from both analytical \cite{Furlanetto:2004ha, hpaseem} and seminumerical/simulation-based \cite{choudhury2021, molaro2019, mesinger2011} approaches.
The SKA will be able to image these ionized bubbles at Cosmic Dawn, and its pathfinder, the Murchinson Widefield Array (MWA; e.g., Ref. \cite{lonsdale2009}), will aim to map the evolution of reionization using the redshifted 21 cm line of neutral hydrogen. The future James Webb Space Telescope (JWST) and European Extremely Large Telescope (ELT) will provide enhanced constraints on the properties of the galaxies responsible for the ionization \cite{park2020, zackrisson2020}. Recently, the Mayall telescope imaged the EGS77 group \cite{tilvi2020} and led to the first observations of ionized bubbles at the highest redshift of $z \sim 7.7$.

There are excellent prospects for investigating reionization using molecular lines. The CO molecular spectrum has a `ladder' of quantum states separated by integer valued quantum numbers, and hence the CO 1-0 line  observations from the epoch of peak star formation, $z \sim 2 - 3$ traced, e.g., by COMAP, also contain a contribution from the CO 2-1 line from the mid to late stages of reionization, $z \sim 6-8$. The latest forecasts predict a detection \cite{breysse2021} of the Reionization signal at high significance
for the next stage COMAP-EoR survey over $z \sim 6 - 8$. It allows us to place very tight constraints on the cosmic molecular
gas density where there is a significant contribution from faint galaxies {\it that would otherwise be missed by current
and future galaxy surveys}, re-iterating the unique ability of line intensity mapping to constrain the properties of the
earliest galaxies. As we saw in Sec. \ref{sec:submmintro}, the redshifted 158 micron line of the singly ionized carbon ion, [CII] and the doubly ionized oxygen ion, [OIII], are also salient probes of reionization and high-redshift galaxies. It can be shown \cite{hpoiii} for a future survey targeting the [OIII] and [CII] species in the sub-millimetre regime at $z \sim 7$ during the period of reionization, which is designed to jointly exploit the potential of the currently proposed EXCLAIM (Experiment for Cryogenic Large-Aperture Intensity Mapping, a balloon based facility), Ref. \cite{cataldo2021} and the FYST (Fred Young Submillimetre Telesope, a ground-based facility \cite{terry2019}) experiments lead to several tens of sigma detection in auto- and cross-correlation modes. Examples of the auto- and cross-correlation power spectra from such a survey at $z \sim 7$ are shown in Fig. \ref{fig:oiii}.

\subsection{Multi-messenger cosmology: the gravitational wave regime}

An outstanding question in cosmology is the formation and fuelling of the earliest black holes in the universe, as we described in Sec. \ref{sec:firstbh}. As we have seen, the bulge mass of the galaxy is connected to that of its central black hole. The results of \cite{behroozi2019} provide a data-driven approach towards constraining the evolution of the stellar mass - halo mass relation as pointed out in Sec. \ref{sec:analytical}. These results indicate that the stellar to halo mass relation evolves only by a factor of $\sim 1.6$ over $z \sim 0-6$, in contrast to the black hole mass - halo mass relation that evolves as a steep function of the halo virial velocity, $M_{\rm BH} \propto v_c^{\gamma}$ where $\gamma \sim 5$.

The above result leads to a fascinating conclusion. Specifically, since the stellar mass evolves much more modestly than the black hole mass (also found in recent observations, e.g., \cite{venemans2016, decarli2018}), the dominance of the black hole can be felt at high redshifts to a much greater distance from the centre of the system. This leads to a 
 direct, observable signature \cite{hploebbh2020} of the prevalence of massive black holes in the centres of the first galaxies during the period of reionization. It is found that the influence of the central black hole can dominate the kinematics  up to a distance of $\gtrsim 0.5$ kpc from the centre of the dark matter halo at redshifts $z \gtrsim 6$.

\begin{figure}
\begin{center}
\includegraphics[width =0.7\textwidth]{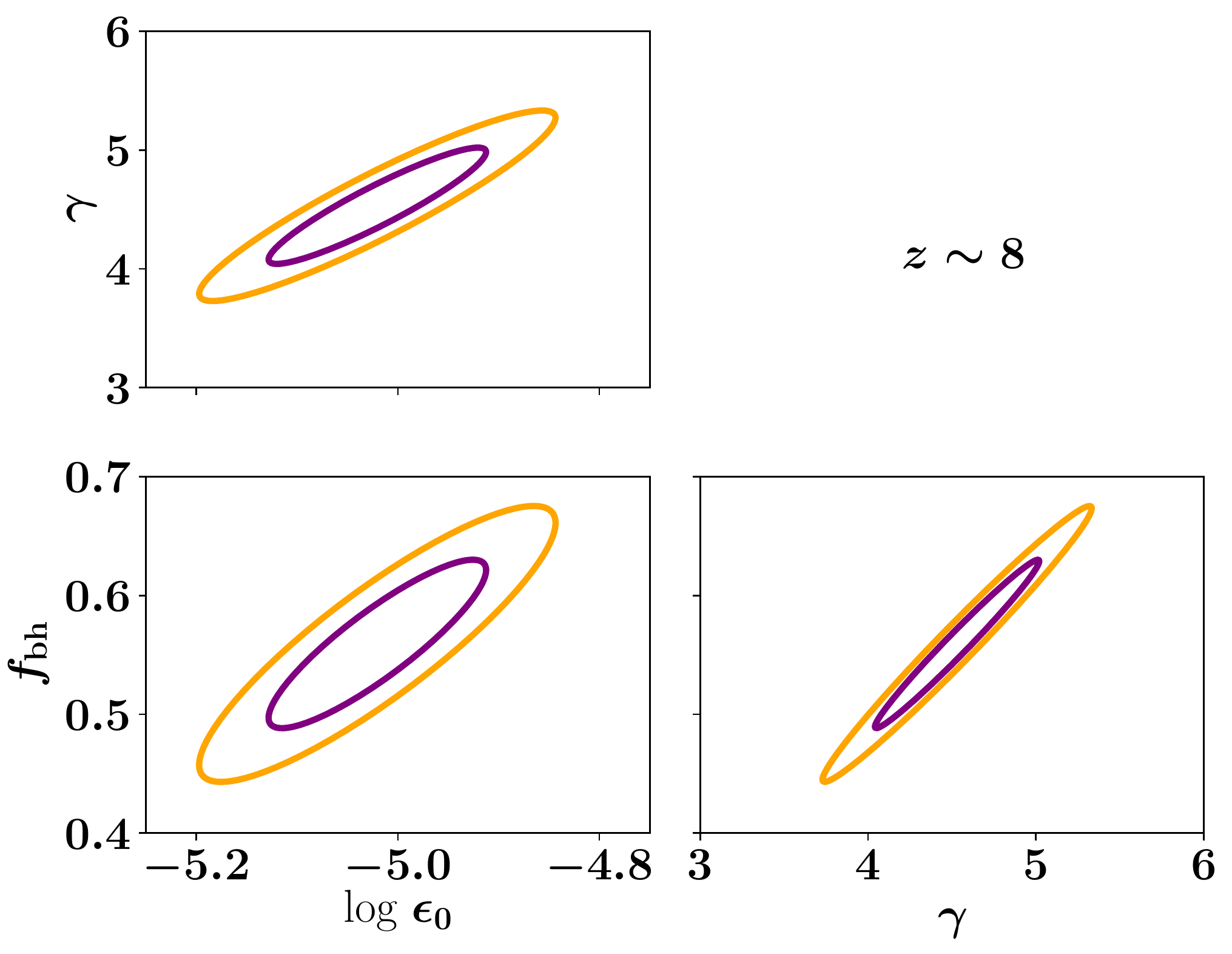}
\end{center}
\label{fig:multimessenger}
\caption{Constraints on the astrophysical parameters in the black hole mass - halo mass relation (the occupation fraction $f_{\rm bh}$, and $\epsilon_0$ and $\gamma$ defined in Table \ref{table:constraints}) from observations of LISA events at $z \sim 8$. Contours indicate 1- and 2-$\sigma$ confidence levels assuming a fiducial 5-year LISA survey having 200 events per year. Figure adapted from \cite{hploeblisa2020c}.}
\end{figure}

Constraining the properties of these earliest sources of reionization, is possible through their gravitational wave signatures detectable by the next-generation Laser Interferometer Space Antenna (LISA) instrument, e.g., \cite{gair2011}. With the above evolution of the black-hole mass to halo mass relation combined with the framework describing merger rates of dark matter haloes, e.g., \cite{fakhouri2010}, it becomes possible to use a future detection rate from LISA to place constraints on the astrophysical parameters (denoted by $f_{\rm bh}$, the occupation fraction, $\epsilon_0$, the normalization and $\gamma$, the slope  described in Sec. \ref{sec:firstbh} and Table \ref{table:constraints}) governing the occupation of black holes in high redshift haloes, similarly to the constraints on astrophysical parameters from baryonic gas occupation of haloes. It can be shown \cite{hploeblisa2020c} that for three different confidence scenarios, each assumed to have 100, 200 or 400 events per year for a survey of a 5 year duration, the above parameters can be constrained to percent or sub-percent accuracy over $z \sim 1-5$ (and out to $z \sim 8$, Fig. \ref{fig:multimessenger}). Even before the advent of LISA, gravitational waves from Pulsar Timing Arrays (PTAs) such as those with the Parkes and the European Pulsar Timing Array (EPTA, \cite{babak2016}) have the potential to provide us with exquisite constraints on the nature of the earliest supermassive black holes. It is possible to identify the potential host galaxies of these systems using LSST on the Vera Rubin Observatory, which is expected to detect $\sim 100-200$ galaxies with black hole masses above $10^{6.5} M_{\odot}$. Considering the gas accretion emission from the merger leads to similar conclusions, reaching about 1-1000 electromagnetic counterparts \cite{kocsis2006}. The above results thus provide a holistic, multi-messenger view into the first galaxies.

We are thus on the brink of significantly advancing our understanding of baryonic cosmology over a nearly 13 billion year timescale, and developing novel techniques that can be extended to other investigations of the early universe. It enables us to combine the largest available datasets of gravitational-wave, radio, millimeter and optical observations to create the most detailed models and simulations of baryonic gas and galaxies, both for the epoch of reionization and the post-reionization universe.

\section{Outlook for the future}
\label{sec:outlook}
In this concluding section, we will summarize some open challenges and recent developments in the areas we have described above, and indicate the theoretical and observational outlook for the future.

\subsection{Unravelling the nature of DLAs}

As we have seen in Sec. \ref{sec:igm}, the intergalactic medium is primarily composed of regions having a column density (in \HI) of $10^{12} - 10^{17}$ cm$^{-2}$.  Above this limit, the hydrogen gas becomes self-shielded to the ionizing radiation, and  
regions that have hydrogen column densities higher than $10^{20.3} \ {\rm cm}^{-2}$ form systems called Damped Lyman Alpha systems (DLAs). These are known to be the highest reservoirs of atomic (hydrogen) gas at intermediate redshifts \cite{wolfe1986, lanzetta1991, storrielombardi2000, gardner1997, prochaska2005} between $z \sim 2$ and $\sim 5$, and our current understanding of the distribution of \HI\ comes from their  observations (Sec. \ref{sec:analytical}) in the spectra of high-redshift quasars \cite{noterdaeme09, noterdaeme12, prochaska09, prochaska2005, zafar2013}.
As the primary reservoirs of neutral gas fuel for the formation of stars at lower redshifts, they are thus the progenitors of today's  galaxies.

Nearly fifty years since they were first discovered \cite{lowrance1972, beaver1972},  a precise understanding of the nature of DLAs still remains elusive. Identifying the host galaxies associated with the absorbers lends clues to their host halo masses and other properties. The presence of the bright background quasar may make the direct imaging of DLAs difficult (though this may also be a function of the impact parameter of observation, e.g., \cite{mackenzie2019}. Imaging surveys for DLAs \cite{fynbo2010, fynbo2011, fynbo2013, bouche2013, rafelski2014, fumagalli2014} thus far point to evidence for DLAs arising in the vicinity of faint, low star-forming galaxies due to the absence of high star-formation rates or luminosities in the samples. There is also an observed lack of high-luminosity galaxies in the vicinity of DLAs, which also points to evidence for DLAs being associated with dwarf galaxies at high redshifts \cite{cooke2015}. The results of simulations find DLAs to arise in host haloes of masses $10^9 - 10^{11} M_{\odot}$ at redshift $z \sim 3$ \cite{pontzen2008, tescari2009, fumagalli2011, cen2012, voort2012, bird2013, rahmati2014}. As far as their structure is concerned, DLAs have been modelled to arise as rotating disks \cite{prochaska2010}, though protogalactic clumps \cite{haehnelt1998} are also consistent with their observed properties. Interestingly, in the redshift regime of relevance to DLAs ($z \sim 2-5$), the statistical halo model framework described in Sec. \ref{sec:dlahimodels} matched to the data assigns an equal likelihood to both these possibilities \cite{hparaa2017}.
 
The cross-correlation of DLAs and the Lyman-$\alpha$ forest from the  twelfth Data Release (DR12) of the Baryon Oscillations Spectroscopic Survey
(BOSS) from the Sloan Digital Sky Survey III (SDSS-III) led to the interesting result \cite{fontribera2012}, that the bias parameter of DLAs -- which is a measure of how strongly the DLAs are clustered  -- is somewhat larger ($b_{\rm DLA} \sim 1.9$) than predicted by the standard evolution expected from lower redshifts, $b_{\rm DLA} \sim 1.5-1.8$ \cite{hptrcar2016}, using the results in \eq{bdla}. Follow-up analyses \cite{perez2018} have measured $b_{\rm DLA} \sim 1.5-2.5$, also finding an increase in the bias with metallicity. 

 The observed high bias $b_{\rm DLA}$ may be consistent with the imaging results if the DLAs arise from dwarf galaxies which are satellites of massive galaxies \cite{fontribera2012}. Theoretically, such a value could arise in models in which the neutral hydrogen in shallow potential wells is depleted, leading to the possibility of very efficient stellar feedback \cite{barnes2014}. However, it is difficult to reconcile models having very strong feedback with the low-redshift observations of \HI\ bias and abundance. 
 The bias value is a crucial pointer to the mass of dark matter haloes hosting the high-$z$ DLA systems, and helps shed light on the (as-yet unsolved) question of the nature of DLAs at high redshifts. It is thus interesting to investigate whether there is scope for placing constraints on the scale- and/or metallicity-dependences \cite{digioia2020} of the clustering using future data. 
 Radio surveys targeting the 21-cm line in \textit{absorption} against bright background quasars (for which the physics follows a description similar to that outlined in Sec. \ref{sec:igm} but does not lead to line saturation), may preferentially pick up larger spiral galaxies and thus be sensitive to more luminous galaxies. The First Large Absorption Survey of Hydrogen (FLASH; Ref. \cite{allison2020}) using the ASKAP telescope, and future surveys with the Canadian Hydrogen Intensity Mapping Experiment \cite{yu2014} aim to look for and localize 21 cm analogs of DLA systems, thus providing a complementary view on the origin of DLAs.

\subsection{Extensions to the halo model framework}
\label{sec:future}

\begin{figure}
\begin{center}
\includegraphics[width = 0.6\textwidth]{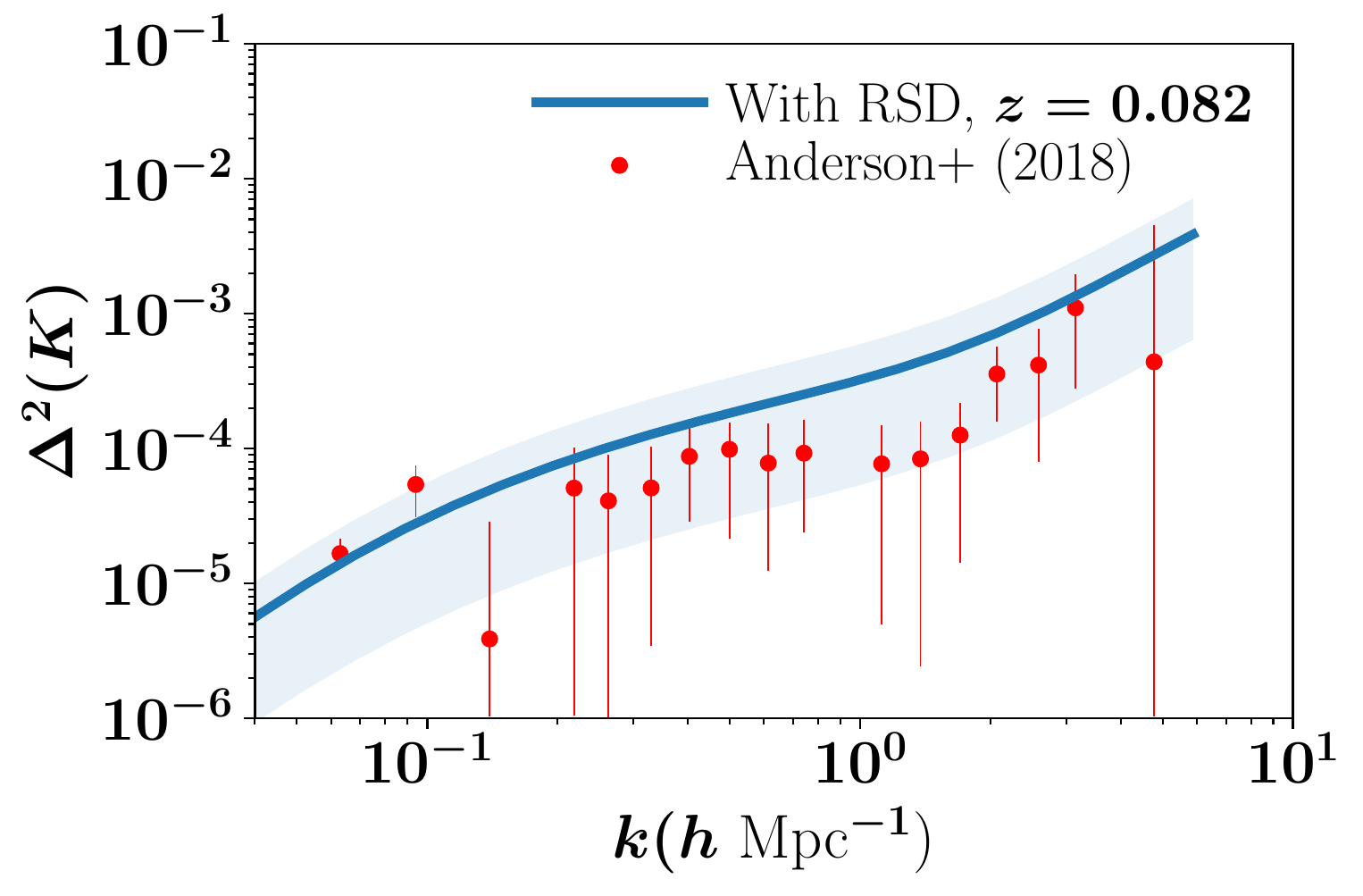}
\caption{Power spectrum in redshift space computed using the halo model framework for HI corrected for the effects of redshift space distortions \eq{rs}. The points with error bars show the measurements of HI-galaxy cross power at $z \sim 0.082$, derived from the cross-correlation of Parkes and 2dF galaxy maps \cite{anderson2018}. The large scale HI bias is  assumed to be $b = 0.85$ with respect to the dark matter, consistently with the findings of \cite{martin12}.}
\end{center}
\label{fig:hiparkeswithrsd}
\end{figure}

Throughout the analysis so far, we have considered the halo model for clustering of dark matter and baryons in the absence of additional non-linear effects like redshift space distortions (RSDs), which originate from the peculiar motions of the gas. It is relatively straightforward to include these effects into the framework when one is interested in analysing real data. Redshift space distortions lead to two main effects on the halo model power spectrum: (i) a boost of power on large scales, due to streaming of matter into overdense regions, and (ii) a suppression of power on small scales due to virial motions within collapsed objects \cite{Kaiser:1987qv, peacock1994}. Both these effects serve to modify the power spectrum in \eq{onehalo} and \eq{twohalo}. By paralleling the approach developed in \cite{seljak2001, white2001} for the galaxy-halo connection, the resulting modifications for the case of gas (using \HI\ as an example) can be shown to have the form :

\begin{eqnarray}
& &P_{\rm HI}(k) = \left( F_{\rm HI}^2+{2\over 3}F_{\rm m}F_{\rm HI} + {1\over 5}F_{\rm m}^2\right)P_{\rm lin}(k)
   \\
& & + 
{1 \over  \bar{\rho_{\rm HI}}^2}
\int n(M) dM {M_{\rm HI}^2}
{\cal R}_2(k\sigma) |u_{\rm HI}(k,M)|^2,
\nonumber
\label{rs}
\end{eqnarray}
where 
\begin{eqnarray}
F_{\rm m}&=&f\int n(M) dM \ b(M) {\cal R}_1(k\sigma) u(k, M) 
\nonumber \\
F_{\rm HI}&=&
{{1}\over \bar{\rho_{\rm HI}}} \int
n(M) d M  M_{\rm HI} (M) b(M) {\cal R}_1(k\sigma)u_{\rm HI} (k,M)
\label{p0}
\end{eqnarray}
where $u(k,M)$ and $u_{\rm HI}(k,M)$ are the normalized Hankel transforms of the dark matter and \HI\ profiles, and
$f = d \log \delta/ d \log a \approx \Omega^{0.6}$ captures the effect of redshift space distortions.
The $\mathcal{R}$ terms are given by [with $\sigma = [GM/2R_v(M)]^{1/2}$ being the rms matter density fluctuation smoothed on the scale of the halo]:
\begin{equation}
  {\cal R}_1(k\sigma) = \sqrt{\pi\over 2} { {\rm erf}(k \sigma/\sqrt{2})\over k \sigma},
\label{r2}
\end{equation}
and
\begin{eqnarray}
  {\cal R}_2(y=k\sigma) &=& {\sqrt{\pi}\over 8} { {\rm erf}(y)\over y^5}
    \left[ 3f^2+4fy^2+4y^4 \right] \nonumber \nonumber \\
  &-& {e^{-y^2}\over 4y^4}\left[ f^2(3+2y^2)+4fy^2 \right]
\end{eqnarray}
\label{r2}
The  above redshift-space modifications can be applied to the \HI\ power spectrum to compare with the recent cross-correlation observations of \cite{anderson2018}, which finds some evidence for a low-amplitude clustering of HI relative to dark matter in the cross-correlation of HI maps from the Parkes survey cross-correlated with galaxies from the 2dF (Two degree Field) survey. The results are shown in Fig. \ref{fig:hiparkeswithrsd}. The observations are well matched to the model within the expected scatter, though there may be some evidence for more significant suppression on scales $k \sim 1$ h Mpc$^{-1}$.

In angular space, the nonlinear modelling of redshift space distortions is more complex and depends on the redshift binning under consideration \cite{jalilvand2020}.
    The treatment of RSDs in the Limber approximation on linear scales has recently been developed \cite{2019MNRAS.489.3385T}, while N-body and hydrodynamical simulations \cite{seehars2016, villaescusa2018} have also been used to model the effect of redshift space distortions on HI intensity maps. Just as in the case of galaxy RSDs, the HI RSDs can potentially constrain several interesting effects and break degeneracies. For example, \cite{hall2017} illustrate how peculiar velocity effects can be constrained 
using the dipole of the redshift space cross-correlation between 21 cm and 
optical surveys.  It is potentially interesting to analyse how redshift space distortions can affect the intensity mapping of submillimetre lines  \cite{chung2019pi} as well, though typically, the large scale effects need a factor $\sim 10$ times more area than available in current configurations, and the small scale effects are often suppressed by the finiteness of the beam in these experiments that wipe out scales of the order of $k \sim 1 - 10 h$ Mpc $^{-1}$ (e.g., Ref. \cite{schaan2021}, as can also be seen from \fig{fig:oiii}). This also makes it difficult to distinguish the 1-halo term from shot noise in these experiments. In the case of cosmological forecasts, these are not expected to make a significant difference to the scales of present interest within the scatter of the results.

\subsection{Observational outlook}
\label{s:results}

One of the most significant challenges for the  detection of a 21 cm signal from the Cosmic Dawn is that of the overwhelming foregrounds. The foreground problem is typically broken into three independent components – Galactic synchrotron, that contributes around 70\% of the total foreground emission \cite{shaver1999}; extragalactic sources which contribute about 27\% \cite{mellema2013}; and Galactic free-free emission which comprises the remaining $\sim$ 1\%. Altogether, these foregrounds are expected to dominate the 21 cm signal brightness temperature by up to 5 orders of magnitude, though this figure reduces to 2–3 when considering the angular brightness fluctuations \cite{bernardi2009}. Furthermore, each source is expected to occupy a different region of angular-spectrum space, e.g., \cite{chapman2016}. A review of recent developments and challenges on the 21 cm data analysis and software fronts is provided in Ref. \cite{liu2020a}.

Current research indicates that foreground removal techniques show a great deal of promise in extracting the 21 cm signal without biases but may introduce larger uncertainties in the measurement of cosmological parameters. This is especially the case for the primordial non-Gaussianity forecasts, since these are degenerate with the large-scale modes which are preferentially washed out by the foregrounds. Recently, it was shown that a multi-parameter fit to the data and foregrounds, taken together, recovers the cosmological parameters extremely well and does not significantly bias the results \cite{fonseca2020}, which thus brings in a very promising outlook for studies of primordial non-Gaussianity. Also, very recently, machine learning tools have been employed to demonstrate the possibility of removing the foregrounds in a satisfactory manner to recover the shapes and sizes of ionised bubbles at the epoch of reionization from SKA and the Hydrogen Epoch of Reionization Array (HERA) images, and found to be robust to instrumental effects \cite{gagnon2021}. It is important to note that foregrounds are not a major problem for line-intensity mapping with other lines -- such as the CO line, which is primarily dominated by point source foregrounds \cite{keating2016} which are about three orders of magnitude larger than the signal, but can be effectively mitigated due to their spectral flatness.

The above main challenge is, however, more than balanced by one of the richest rewards in the domain of cosmology and astrophysics: to be able to capture, for the first time, the evolution of the Universe almost up to the beginning of the dark ages, closing the gap between the local universe and the CMB, approaching an enormous amount of information contained in nearly $10^{15}$ independent Fourier modes, representing the largest cosmological dataset.
 On a broader scale, the data science and machine learning tools needed for achieving the goals are also at the forefront of several research directions today.

\subsection{Fundamental physics from the Cosmic Dawn}\label{s:applications}

Including redshifts up to the Cosmic Dawn in a fully data-driven formalism of baryonic gas and galaxies opens up the exciting possibility of constraining Fundamental Physics from the best combination of the data we have. Most importantly, it facilitates realistically addressing these questions, for the first time, from a thorough understanding of the astrophysics. This is crucial since an inadequate modelling of the astrophysics in surveys can indicate false inconsistencies in different data sets and lead us to incorrect conclusions, like e.g., new physics. Executing this objective would thus be the most ambitious and groundbreaking theme of exploration in the coming years.  \cite{weltman2020} explores ways in which the SKA will deliver unprecedented constraints that can transform our understanding of fundamental physics.  

At low redshifts, intensity mapping has already been shown to be a powerful probe of dark energy \cite{chang10}; the acoustic oscillations in these maps may constrain dark energy out to high redshifts \cite{wyithe2008a}. 21 cm cosmology allows a unique test of the (possible) variation of fundamental constants (such as the fine structure constant and electron mass) across space and time, due to its ability to probe a huge range of redshifts up to Cosmic Dawn ($z \sim 30$). In particular, it was shown that such variations are constrainable at the level of 1 part in 1000 with the upcoming SKA \cite{lopez2020}. The recent EDGES signal of 21-cm at Cosmic Dawn has been interpreted as an indication of dark matter being possibly composed of axions \cite{sikivie2019}; in general, the imprint of axion dark matter on the epoch of reionization has been explored in \cite{carucci2019}. Another interesting prospect for fundamental physics from the Cosmic Dawn is to constrain the running of the spectral index of the matter power spectrum with an interferometer (of size 300 km) probing up to the dark ages\cite{weltman2020}.

There are signs to indicate that the primordial gravitational wave background can place constraints on non-Gaussianity at early times using future interferometers such as LISA and the SKA \cite{unal2019}. In \cite{cai2019}, it was found that if dark matter consists of primordial black holes with asteroid masses, the accompanying gravitational waves occur at mHz frequencies and are detectable by LISA irrespective of the local non-Gaussianity. It is also possible to place limits on the fraction of dark matter in different sub-species, such as axions \cite{unal2020}, by using the spin of black holes in Active Galactic Nuclei which emit radiation in the [OIII] forbidden line transition. Similarly, inflationary gravitational waves can be probed through the circular polarisation of the 21-cm line (e.g., \cite{mishra2018}). An important use of the gravitational wave measurements from standard sirens is to constrain the value of the Hubble parameter and its evolution with cosmic time, shedding light into the presently unresolved problem of the discrepancy between the local and CMB measurements of the Hubble parameter (e.g., \cite{Riess_2019}).

The intrinsic dipole of the large-scale structure distribution, as traced by galaxies, and its potential deviation from that of the CMB is an important test of the isotropy of the universe, one of the cornerstones of the standard cosmological model. Recent work \cite{nadolny2021} shows that it may be possible to measure this dipole independently of the dipole associated with the speed and direction of our motion,  using a large catalog of radio sources such as measured with a SKA-like radio survey in the coming years. This is an important probe of fundamental physics which can be refined with future large datasets.

The signatures of the first black holes in the gravitational wave, millimetre and radio wavebands can be used to combine this with an understanding of the co-evolution of the first black holes and galaxies at the epoch of Cosmic Dawn. This will also provide insights into a hitherto unresolved problem in cosmology: how did the first black holes form, and what was their contribution to reionization?

Finally, the baryonic gas and galaxy parametrizations, together with the latest theoretical insights into the formation of the first structures at Cosmic Dawn, can be used to investigate the physics constraints on the nature of dark matter, dark energy, theories of gravity and the Universe’s earliest moments, that one can glean from the largest fraction of the Universe observed till date. There is already tremendous promise in the fact that wider surveys (such as the CHIME and the SKA) are able to achieve much tighter constraints on cosmological parameters, even fully alleviating the degradation caused by the astrophysics. With the advances in theoretical understanding of physics at its most extreme — such as close to the first supermassive black holes — we can be expected to gain insights into constraining a quantum theory of gravity.

\section{Acknowledgments}
I acknowledge support from the Swiss National Science Foundation (SNSF) via Ambizione grant PZ00P2\_179934.

\bibliographystyle{unsrt}

\def\aj{AJ}                   
\def\araa{ARA\&A}             
\def\apj{ApJ}                 
\def\apjl{ApJ}                
\def\apjs{ApJS}               
\def\ao{Appl.Optics}          
\def\apss{Ap\&SS}             
\def\aap{A\&A}                
\def\aapr{A\&A~Rev.}          
\def\aaps{A\&AS}              
\def\azh{AZh}                 
\def\baas{BAAS}
\def\jcap{JCAP}
\def\jrasc{JRASC}             
\def\memras{MmRAS}
\def\na{New Astronomy}
\def\nat{Nature}
\def\mnras{MNRAS}             
\def\pra{Phys.Rev.A}          
\def\prb{Phys.Rev.B}          
\def\prc{Phys.Rev.C}          
\def\prd{Phys.Rev.D}          
\def\prl{Phys.Rev.Lett}       
\def\pasp{PASP}    
\def\pasa{PASA}           
\def\pasj{PASJ}
\def\physrep{Phys. Repts.}
\def\qjras{QJRAS}             
\def\skytel{S\&T}             
\def\solphys{Solar~Phys.}     
\def\sovast{Soviet~Ast.}      
\def\ssr{Space~Sci.Rev.}      
\def\zap{ZAp}                 
\let\astap=\aap
\let\apjlett=\apjl
\let\apjsupp=\apjs

\bibliography{mybib, refs, mybib1, mybib1a, mybib2, mybib3, mybib4, mybib5, mybib6, main_bib}

 \end{document}